\documentclass[amsmath,amssymb,nofootinbib]{revtex4}
\usepackage[english]{babel}
\usepackage[dvips]{graphicx}

\begin{document}

\title{Black Holes as Dark Matter Annihilation Boosters}
\author{Mattia Fornasa$^{1,2}$}
\author{Gianfranco Bertone$^2$}
\affiliation{$^1$ INFN, Sezione di Padova, via Marzolo 8, Padova, Italy}
\affiliation{$^2$ Institut d'Astrophysique de Paris, UMR 7095-CNRS, Universit\'e Pierre et Marie Curie, 98 bis Boulevard Arago 75014, Paris, France}
\email{mfornasa@pd.infn.it,gianfranco.bertone@iap.fr}

\begin{abstract}
We review the consequences of the growth and evolution of Black Holes 
on the distribution of stars and Dark Matter (DM) around 
them. We focus in particular on Supermassive and Intermediate-Mass 
Black Holes, and discuss under what circumstances they
can lead to significant overdensities in the surrounding 
distribution of DM, thus effectively acting as DM Annihilation 
Boosters. 
\end{abstract}

\maketitle

\tableofcontents

\section{Introduction}
\label{sec:one}

To identify the nature of Dark Matter (DM) particles, three different 
experimental strategies have been devised. First, high-energy colliders like
the Large Hadron Collider at CERN \cite{LHC} may soon be able to produce
DM particles, and possibly identify them (see e.g. 
Refs. \cite{Baltz_Battaglia_Peskin_Wizanski} and 
\cite{Datta_Kong_Matchev} and references therein). Another possibility is
{\it Direct Detection}, based on the search for rare events in which a DM 
particle scatters off the nuclei of large detectors (see e.g. 
Ref. \cite{Munoz} for a review). Finally, {\it Indirect Detection} is based 
on the search for the products of annihilation or decay of DM particles (such 
as photons, neutrinos and anti-protons). 
Many candidates have been proposed for DM, but the most studied ones belong 
to the class of Weakly Interacting Massive Particles (WIMPs) 
\cite{Bertone_Hooper_Silk, Bergstrom}, that includes, among others, 
the lightest neutralino in Minimal or Non-Minimal Supersymmetric 
extensions of the Standard Model \cite{Jungmann_Kamionkowski_Griest}, 
and the lightest Kaluka-Klein state in models with Unified Extra Dimensions 
\cite{Appelquist_Cheng_Dobrescu,Servant_Tait,Cheng_Matchev_Schmaltz}.

Here we focus on Indirect DM searches. Ideal targets for Indirect DM searches
are nearby, high DM density regions, such as local celestial bodies like the
Earth and the Sun, where DM can be captured, or centers of galactic halos
(see e.g. Refs. 
\cite{Aharonian_et_al,Zaharijas_Hooper,Profumo,Mambrini_Munoz_Nezri_Prada,Bertone_Merritt_two,Cesarini_Fucito_Lionetto_Morselli_Ullio,Bouquet_Salati_Silk}).
Two classes of astrophysical objects of particular interest are SuperMassive 
Black Holes (SMBHs) \cite{Ferrarese_Ford,Kormendy_Ho}, with masses 
from $ 10^6 $ to $ 10^9 M_{\odot} $ and the more speculative Intermediate 
Mass Black Holes (IMBHs), with a mass from $ 20 M_{\odot} $ to 
$ 10^6 M_{\odot} $ (see e.g. Refs. \cite{Miller_Colbert,Koushiappas_Bullock_Dekel,Bertone_Zentner_Silk} and 
references therein). Both these classes of compact objects can influence the 
distribution of DM in which they are embedded, leading to strong 
overdensities: we review here the impact of the formation and growth of BHs 
on the surrounding distribution of matter, and the consequences for Indirect 
DM searches.
Being the annihilation flux proportional to the integral of the DM density
squared, scenarios where the density is {\it boosted} by the presence, or 
the growth, of a central BH, are very promising for Indirect searches.
BHs can thus be considered as {\it DM Annihilation Boosters}.

The paper is organized as follows: in Section \ref{sec:two} we discuss our
current understanding of DM profiles in halos without central BHs. In Section
\ref{sec:three} we address the case of a DM distributon evolving around an
already-formed BH. In Section \ref{sec:four}, we discuss the case of 
adiabatic growth and the conditions under which DM ``spikes'' can form.
In Section \ref{sec:five} we will focus on IMBHs, and show that some of the 
problems associated to the formation of spikes around SMBHs can be evaded,
with important consequences for Indirect DM searches. Conclusions will be 
drawn in Section VI.

\section{Dark Matter profiles without Black Holes}
\label{sec:two}
There is strong evidence in favor of the presence of SMBHs at the center of
every galaxy with a substantial bulge component \cite{Ferrarese_Ford,Kormendy_Ho,Merritt} and it has been suggested that even 
globular clusters can harbor IMBHs \cite{Miller_Colbert}.
Throughout the paper, we will generically refer to large gravitationally 
bound systems like globular clusters, galaxies and clusters of galaxies, as
{\it galaxies}, and we will refer to their central region as the 
{\it nucleus}, which may host BHs. Such compact objects account roughly for 
$ 10^{-3} \% $ of the baryonic mass of the galaxy, which is composed of stars, 
intergalactic dust and DM. 
We focus on a generic WIMP scenario, with an annihilation cross 
section of order $ \sigma v \thickapprox 10^{-26} \mbox{cm}^3 \mbox{s}^{-1} $ 
and a mass ranging between the GeV and the TeV scale.

Since we want to characterize how BHs influence the surrounding distribution 
of matter, we need to specify how DM is distributed {\it before}
the BHs form, and use this information as initial condition for the problem
at hand. Profiles without any central object also receive particular 
attention {\it per se}, since the cuspiness of a DM halo without BH can give 
informations about the ``coldness'' of the DM candidate \cite{Tremaine_Gunn}. 
Recently, $ N $-body simulations of galaxies found results in favor 
of power law profiles (with slope from $ -1 $ to $ -1.5 $) for the nuclear 
region, emphasizing the contrast with direct observations, such as rotation 
curves of Low Surface Brightness galaxies (LSBs) \cite{deBlok_Bosma,deBlok,Gentile_Burkert_Salucci_Klein_Walter} and X-ray imaging, 
which suggest instead the presence of flat DM cores. 

Navarro {\it et al.} \cite{Navarro_Hayashi_Power_Jenkins_et_al} and Reed 
{\it et al.} \cite{Reed_Governato_Verde_Gardner_et_al} used $ N $-body 
techniques to simulate the high resolution evolution of galaxies with masses 
that go from dwarf galaxies ($ 10^{10} M_{\odot} $) to clusters of galaxies 
($ 10^{15} M_{\odot} $). They fitted the final density profiles with typical 
parametrizations proposed by Navarro, Frenk and White
\cite{Navarro_Frenk_White} (NFW):
\begin{equation}
\rho(r)=\frac{\rho_0}{r/r_s(1+r/r_s)^2},
\label{eqn:NFW}
\end{equation}
and by Moore \cite{Moore_Quinn_Governato_Stadel_et_al} (M99):
 \begin{equation}
\rho(r)=\frac{\rho_0}{(r/r_s)^{1.5}(1+r/r_s)^{1.5}},
\label{eqn:M99}
\end{equation}

finding that simulated data are well approximated by such profiles, 
that are, hence, ``universal'', in the sense that the same analytical form
successfully captures the shape of halos at different masses.
However, the logarithmic slope $ \beta(r)=d\ln \rho(r)/d \ln r $ of the 
density profile decreases faster (more slowly) in the simulated data than 
does in the NFW (M99) profile at small radii. 
Moreover, $ N $-body simulations do not exhibit any indications that 
$ \beta(r) $ converges to a central value $ \beta_0 $, as should happen for 
a NFW profile ($ \beta_0=-1.0 $) or for a M99 profile ($ \beta_0=-1.5 $). This 
can be due to the finite resolution of numerical simulations, which can be 
trusted down to the resolution radius $ r_{min} $, usually taken to be around 
0.5\% of the virial radius, depending on the total number of particles in the
simulation ($ r_{min} \thickapprox \mbox{ kpc} $, for Milky Way-sized halos). 
The structure of the inner region therefore is not clear yet, and the value 
of the central slope $ \beta_0 $ can only be inferred by extrapolation. 
Since the region near $ r_{min} $ is where the deviations from 
Eqs. \ref{eqn:NFW} and \ref{eqn:M99} are stronger, the extrapolation procedure
can lead to significant errors.

Aside from these uncertaintes, Navarro {\it et al.}
\cite{Navarro_Hayashi_Power_Jenkins_et_al} used the density of their simulated
halos to state that, given a generic power law profile whose central slope
converge to a finite value, such value cannot be steeper than $ -1.5 $. 
Throughout this paper, we will consider this value as the lower limit for the 
slope of a DM nuclear profile in absence of BH.
The profiles in Eqs. \ref{eqn:NFW} and \ref{eqn:M99} are particular cases of
a more general parametrization, the so-called $ (\alpha,\beta,\gamma) $
profile \cite{Zhao}:
\begin{equation}
\rho(r)=\rho_0 \left( \frac{r_0}{r} \right)^{\gamma}
\frac{1}{[1+(r/r_0)^{\alpha}]^{(\beta-\gamma)/\alpha}},
\end{equation}
that reduces to $ \rho \propto r^{-\gamma} $ 
($ \rho \propto r^{-\beta} $) in the limit of small (large) radii, while $ \alpha $ 
characterizes the sharpness of the change in the logarithmic slope. The NFW is 
recovered for $(\alpha,\beta,\gamma) = (1,3,1) $ and the M99 
for $ (\alpha,\beta,\gamma)=(1,3,1.5) $.

It has been proposed that, for a choice of parameters mimicking the presence
of flat cores (e.g. if $ (\alpha,\beta,\gamma)=(2,3,0.2) $), such profiles
provide good fits of both the rotation curves from direct observation of LSB 
galaxies and of the rotation curves from simulated nuclei, suggesting that 
core profiles should be preferred to Eqs. \ref{eqn:NFW} and \ref{eqn:M99} 
since no contrast between $ N $-body techniques and observational data is 
present \cite{Kravtsov_Klypin_Bullock_Primack}. A possible explanation for
the different results obtained by Navarro {\it et al.} 
\cite{Navarro_Hayashi_Power_Jenkins_et_al} and Kravtsov {\it et al.}
\cite{Kravtsov_Klypin_Bullock_Primack} is the fact that the analysis by
Navarro {\it et al.} does not take into account the large uncertainty
in the determination of the luminosity distance of galaxies.

Moreover, it was recently  shown \cite{Merritt_Navarro_Ludlow_Jenkins} that 
the best fit to simulated data for high-resolution $ \Lambda $CDM halos is 
obtained with profiles inspired from the so-called S\'ersic law \cite{Sersic}:
\begin{equation}
\ln(\Sigma/\Sigma_e)=-b(X^{1/n}-1).
\label{eqn:Sersic}
\end{equation}

Such a relation provides the best description of the luminosity profiles of
elliptical galaxies and the bulges of disk galaxies \cite{Graham_Guzman}: 
$ \Sigma $ is the projected density, $ X=R/R_e $, and $ R $ is the projected 
radius. 
The parameter $ n $, called S\'ersic index, defines the shape of the profile, 
and $ b $ is a function of $ n $, usually chosen so that the radius $ R_e $
contains half of the luminosity of the galaxy.

Eq. \ref{eqn:Sersic} can be re-written as
\begin{equation}
\frac{d \ln \Sigma}{d \ln R}=-\frac{b}{n} \left( \frac{R}{R_e} \right)^{1/n},
\end{equation}
making explicit the power law behaviour of the logarithm slope. Parametrizing 
the spatial DM profile in a similar way, we obtain
\begin{equation}
\frac{d \ln \rho(r)}{d \ln r}=-2 \left( \frac{r}{r_{-2}} \right)^{1/n}
\label{eqn:Einasto}
\end{equation}
or $ \rho(r) \thickapprox \exp(-Ar^{1/n}) $, that is called {\it Einasto 
profile}, in order to emphasize the difference from the S\'ersic law, since we 
are using now spatial, not projected, quantities. $ r_{-2} $ is the distance
from the center where the slope is equal to $ -2 $. Eq. \ref{eqn:Einasto} was 
tested fitting the density of the DM halos simulated by Navarro {\it et al.} 
\cite{Navarro_Hayashi_Power_Jenkins_et_al}, providing better results than a 
NFW profile, at least for dwarf and galaxy halos 
\cite{Merritt_Navarro_Ludlow_Jenkins}. The values of $ n $ (now called the 
Einasto index), left as a free parameter in the fit, range from 4.33 to 7.44. 
A more detailed description of this model can be found in Merritt {\it et al.} 
\cite{Merritt_Graham_Moore_Diemand_Torzic}, where fits
to simulated DM halos are studied and compared with alternative 
parametrizations. A number of recent studies \cite{Gao_Navarro_Cole_et_al,Hayashi_White} have confirmed that simulated DM density profiles deviate 
slightly but systematically from the NFW form and are better approximated by 
Einasto's empirical law.

If the Einasto relation was confirmed as a good parametrization of the inner 
region of DM halos, this would suggest that a scale-free 
relation like Eq. \ref{eqn:Einasto}, describing both dark and luminous matter, 
is a characteristic feature for systems that form via gravitational clustering.

\section{Particle density around already-formed Black Holes}
\label{sec:three}
\subsection{The Fokker-Planck equation and the Bahcall-Wolf solution}
\label{sec:Bahcall_Wolf}

A population of particles (both stars and DM particles) around a BH can be 
described by a distribution function $ f({\bf x},{\bf v},t) $, whose evolution
is governed by gravitational encounters among particles \cite{Spitzer}. In the
small-angle approximation, such distribution function slowly diffuses in the 
phase space $ ({\bf x}, {\bf v}) $ towards a steady-state configuration.
The time needed to achieve this equilibrium solution is defined as the
relaxation time $ t_{rel} $. For the stellar population, assuming that all 
stars have the same mass $ m_{\star} $ \cite{Spitzer}:
\begin{eqnarray}
\label{eqn:relaxation}
t_{rel} & \thickapprox & \frac{0.34 \sigma^3}{G^2 \rho m_{\star} \ln\Lambda} \\
& \thickapprox & 0.95 \cdot 10^{10} \mbox{ yrs} \left( \frac{\sigma}{200 
\mbox{ km s}^{-1}} \right)^3 \left( \frac{\rho}{10^6 M_{\odot} \mbox{pc}^{-3}} 
\right)^{-1} \left( \frac{m_{\star}}{M_{\odot}} \right)^{-1} \left( 
\frac{\ln\Lambda}{15} \right)^{-1}, \nonumber
\end{eqnarray}
where
$ \sigma $ is the velocity dispersion, $ \rho $ the stellar density and 
$ \ln\Lambda $, known as the Coulomb logarithm, comes from imposing a physical
upper cut-off in the distribution of impact parameters for stellar encounters.
$ \ln\Lambda $ is usually related to the mass of the central BH 
($ M_{\bullet}$) expressed in units of stellar masses 
\cite{Preto_Merritt_Spurzem}: 
\begin{equation}
\ln\Lambda \thickapprox \ln \left( \frac{r_h\sigma^2}{2Gm_{\star}} \right)= 
\ln \left( \frac{M_{\bullet}}{2m_{\star}} \right)=\ln(N_{\bullet}/2).
\label{eqn:Coulomb_logarithm}
\end{equation}

$ t_{rel} $ depends on the distance from the center of the galaxy but usually, 
as in Eqs. \ref{eqn:relaxation} and \ref{eqn:Coulomb_logarithm}, it is 
computed at the influence radius $ r_h $, defined as the radius at which the 
gravitational potential due to the BH is equal to the kinetic energy:
\begin{equation}
r_h=\frac{GM_{\bullet}}{\sigma^2} \thickapprox 11 \mbox{ pc} \left( 
\frac{M_{\bullet}}{10^8 M_{\odot}} \right) \left( \frac{\sigma}{200 
\mbox{ km s}^{-1}} \right)^{-2}.
\label{eqn:influence_radius}
\end{equation}

In the case a singular isothermal density profile \cite{Spitzer} 
($ \rho(r)=\sigma /2 \pi Gr^2 $), $ M(r \leq r_h)=2M_{\bullet} $; and this can 
be used as an alternative definition of the influence radius. For the Milky 
Way (MW), $ r_h \thickapprox 3 \mbox{ pc} $ according to both definitions.
There seems to be a clear trend of relaxation times with the mass of the 
central BH, where smaller objects (corresponding to fainter nuclei) are 
associated to smaller relaxation times. It can be seen from 
Fig. 1 (taken from Ref. \cite{Merritt}) using the empirical 
$ M_{\bullet}-\sigma $ relation \cite{Ferrarese_Ford}:
\begin{equation}
 M_{\bullet}=5.72 \cdot 10^6 M_{\odot}
\left( \frac{\sigma}{10 \mbox{km s}^{-1}} \right)^{4.86}.
\label{eqn:M_sigma}
\end{equation}
Nuclei can then be classified in two different categories. Those nuclei with a
relaxation time larger that the Hubble time, cannot have already achieved 
their relaxed equilibrium configuration, so that their distribution will 
reflect the process of nuclear formation. They are called 
{\it collisionless nuclei}, and they are characterized by a central region 
with a low density of stars, since near the BH a core is present with 
a slope $ \lesssim 0.2 $ \cite{Merritt}, at least for those nuclei where the 
influence radius is resolved. The ``mass deficit'' (compared to what one 
expects from the S\'ersic law \cite{Sersic}) is up to 4 times the mass of the 
central BH. There are, then, galaxies, like the MW and M32 that are 
characterized by a relaxation time smaller than $ 10^{10} \mbox{ yrs} $ 
(at resolved radii, e.g. the MW has $ t_{rel}=3.5 \times 10^9 \mbox{ yrs} $ 
at radius $ \thickapprox 0.1 \mbox{ } r_h $). These {\it collisional nuclei} 
have already reached their steady-state configuration.  Usually they are faint nuclei 
($ M_V \lesssim -20 $) and, opposite to cores of collisionless nuclei, the 
innermost region exceedes the S\'ersic law, establishing an inner power law 
profile with slope steeper than $ \thickapprox 1.5 $ or a compact stellar
nucleus \cite{Cote_et_al_one}.

\begin{figure}[!tb]
\begin{center}
\label{fig:relaxation_time}
\includegraphics[width=0.6\textwidth]{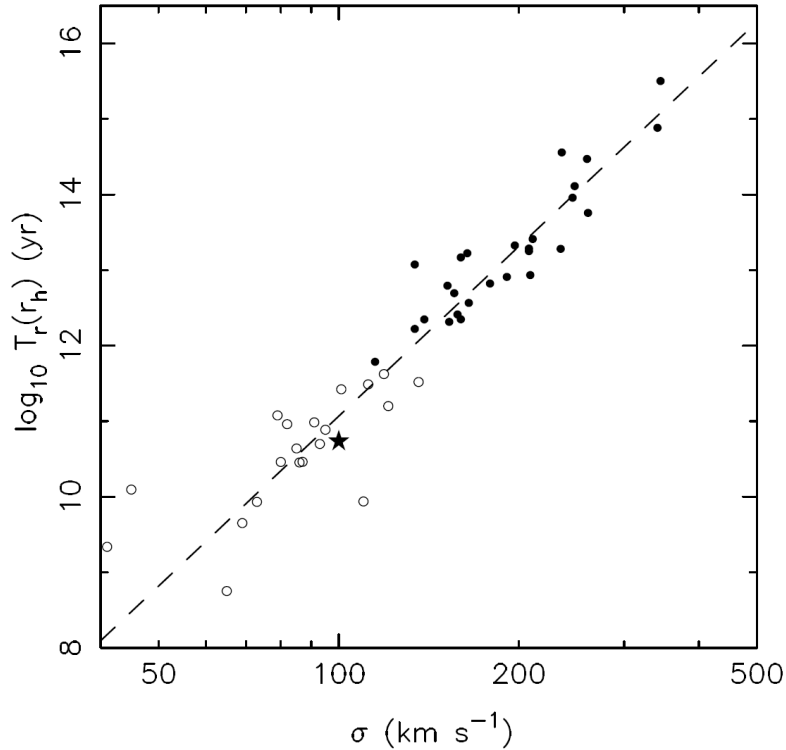}
\caption{Relaxation times measured at the SMBH influence radius in the ACS/Virgo sample of galaxies (see also C\^ot\`e {\it et al.} (2004)), versus the central stellar velocity dispersion. Filled symbols are nuclei in which the influenced radius is resolved. The star is the MW. Figure taken from Merritt (2006).}
\end{center}
\end{figure}

In this Section we will restrict ourselves to the case of collisional nuclei
where the relaxation time is smaller than the Hubble time and the nucleus
has, today, a relaxed, steady-state equilibrium configuration for the 
stellar population.
The diffuse evolution of an isotropic distribution $ f $ is described by the 
Fokker-Planck equation \cite{Spitzer,Preto_Merritt_Spurzem}, where 
gravitational collisions are taken into account and parametrized as:
\begin{equation}
4\pi^2 p(E)\frac{\partial f}{\partial t} = -\frac{\partial F_E}{\partial E}=
\frac{\partial}{\partial E} \left[ -D_{EE}\frac{\partial f}{\partial E}-D_E f
\right],
\label{eqn:Fokker_Planck}
\end{equation}
with
\begin{equation}
D_{EE}(E)=64\pi^4 G^2 m^2 \ln\Lambda \left[ q(E) \int_{-\infty}^{E} 
dE^{\prime}f(E^{\prime}) + \int_{E}^{0} dE^{\prime} q(E^{\prime}) f(E^{\prime})
\right],
\end{equation}
\begin{equation}
D_E(E)=-64\pi^4 G^2 m^2 \ln\Lambda \int_{E}^{0} dE^{\prime} 
p(E^{\prime}) f(E^{\prime}),
\label{eqn:D_E}
\end{equation}
\begin{equation}
q(E,t)=\frac{1}{3}\int_0^{r_{max}} v^3 r^2 dr= \frac{1}{3} \int_0^{r_{max}}
[2(E-\phi)]^{3/2} r^2 dr,
\end{equation}
while $ p(E)=-\partial q/\partial E $ is the volume of phase-space accessible 
to stars with energy $ E $.

The equilibrium solution cannot be a Maxwellian distribution, since it
would imply an unphysical stellar density near the BH 
\cite{Shapiro_Teukolski}, given that stars cannot be present at radii smaller 
than the tidal radius $ r_t $, inside which tidal forces tear stars apart. 
As a consequence, the distribution function is set to zero for $ r \leq r_t $.

The physical steady-state solution was determined by Bahcall and Wolf 
\cite{Bahcall_Wolf}, following a previous work of Peebles \cite{Peebles}.
They proposed that the equilibrium configuration is a zero-flux solution, 
and obtained a distribution function with a power-law behaviour 
$ f(E) \propto |E|^{1/4} $, with a corresponding power-law density profile 
$ \rho(r)=\rho_0 r^{-7/4} $. They also numerically solved a Fokker-Planck-like
equation, obtaining a profile that can be very well described by the
zero-flux solution, in the inner region ($ r \lesssim 0.2 \mbox{ } r_h $, 
where the cusp actually forms), and a Keplerian rise in the velocity 
dispersion $ \sigma \propto r^{-1/2} $. See Fig. 2 taken from 
Ref. \cite{Merritt}.
Their solution has been confirmed by $ N $-body simulations 
\cite{Preto_Merritt_Spurzem}, in which the assumptions of isotropy
and small-angle, characteristics of the Fokker-Planck-like formalism,
had been relaxed.

\begin{figure}[tbh]
\begin{center}
\label{fig:FP_Nbody}
\includegraphics[width=0.8\textwidth]{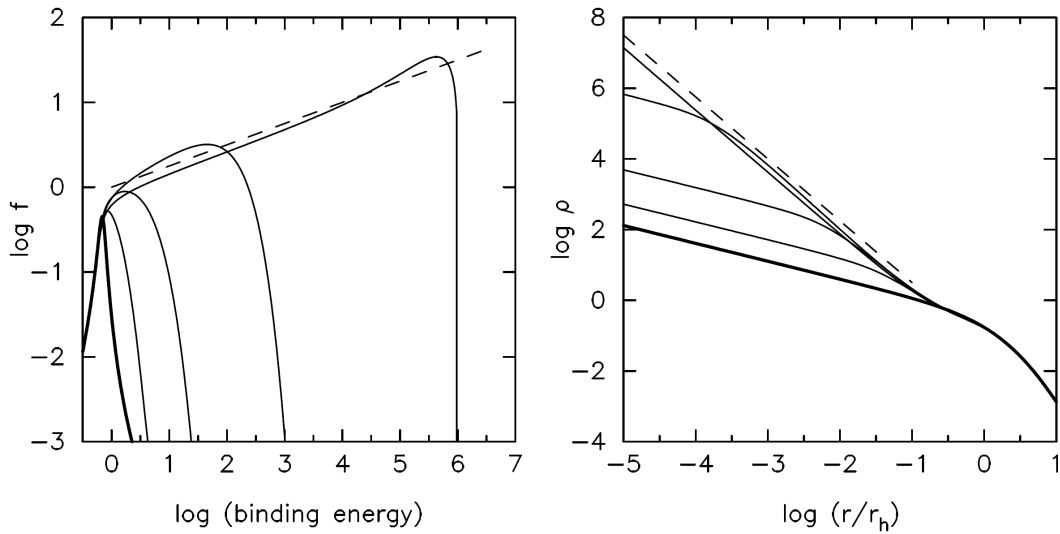}
\caption{Evolution of the stellar distribution around a SMBH due to energy exchange between stars. These curves were computed from the isotropic, orbit-average Fokker-Planck equation with boundary condition $ f=0 $ at $ \log|E|=6 $. {\it Left panel:} phase-space density $ f $; {\it right panel:} configuration-space density $ \rho $. The initial distribution (shown in bold) had $ \rho \propto r^{-0.5} $ near the SMBH; thin curves show $ f $ and $ \rho $ at times of $ (0.2, \mbox{ } 0.4, \mbox{ } 0.6, \mbox{ } 1.0) $ in units of the relaxation time at the SMBH's initial influence radius $ r_h $. Dashed line show the ``zero-flux'' solution $ f \propto |E|^{1/4} $ and $ \rho \propto r^{-7/4} $. The steady-state density is well approximated by the zero-flux solution at $ r \lesssim 0.2 \mbox{ } r_h $. Figure taken from Merritt (2006).}
\end{center}
\end{figure}

The validity of Eq. \ref{eqn:Fokker_Planck} relies on the following 
assumptions:
\begin{itemize}
\item stars are point-like masses, described by a distribution function that 
evolves due to gravitational interactions with the central BH and among 
themselves. In particular, this means that encounters in which stars collide 
with each other are neglected
\item the small-angle approximation: the gravitational potential belongs to a 
particular class of interactions for which the net force experienced by a 
test particle surrounded by a population of other bodies with which it 
interacts, is mainly due to cumulative, weak encounters with particles far 
away, rather than to strong but infrequent interactions with close particles.
If we call $ p_0 $ the impact parameter for the interaction that causes a 
deflection of $ 45^{\circ} $ in the velocity of the test particle, all the 
close encounters with impact parameters $ p \leq p_0 $ count only for  
4\% of the total net deflection \cite{Spitzer}.
Working under the small-angle approximation means that, in the study of the
evolution of a test particle with a velocity {\bf v} embedded in a larger 
particle population, we are considering only the encounters with those bodies 
far away that will produce small deflections  $ {\bf \Delta v} $ compared to 
the initial velocity {\bf v}
\item the distribution function does not depends on the angular momentum
and, since the gravitational potential is function only of the radial 
coordinate, the stellar distribution is isotropic
\item the mass of the central BH is much larger than the mass of a star
and does not change with time, so that inside the influence radius, the 
gravitational potential is constant and Keplerian and is due only to the BH 
itself. Under this assumption we can (as in Eq. \ref{eqn:Fokker_Planck}) 
neglect the term proportional to $ \partial f/ \partial E $ that would appear 
in a more complete form of the Fokker-Planck equation.
Moreover the BH mass has to be much smaller than the total stellar mass near 
the BH itself. Requiring this particular mass hierarchy \cite{Bahcall_Wolf},
leads us to a consequent timescale hierarchy, i.e., the assumption
$ m_{\star} \ll M_{\bullet} \ll M_{\star}(r<r_h) $ implies that the crossing 
time $ t_{cr} $ (the time needed for a star to cross the nucleus) is much 
shorter than the relaxation time, so that in a crossing time stars do not 
experience any changes in the physical proprieties of the system.
\end{itemize}

All the assumptions listed above can also be satisfied in the case of a 
distribution of DM particles only (with a common mass $ m_{\chi} $), so that a 
Fokker-Planck formalism is appropriate also for DM. We expect the existence of 
a relaxed solution also in this case, but the relaxation timescale for DM is 
enormously larger than for stars and DM particles will never reach their 
steady-state solution: they can be effectively considered as collisionless 
objects, pratically not sensitive to the gravitational self-interactions that 
drive the dynamical evolution described by the Fokker-Planck equation.

The requirement of a common mass for particles (stars or DM) is not included 
in the list above, because in the more realistic case of a nucleus with
particles of different mass, Eq. \ref{eqn:Fokker_Planck} can be modified 
in order to describe a multi-mass case. For a two-component nucleus, made of 
stars (with a common mass $ m_{\star} $) and DM particles with 
$ m_{\chi} \ll m_{\star} $ the Fokker-Planck equations will be the following:
\begin{equation}
4\pi^2 p(E) \frac{\partial g_{\star}}{\partial t}=\frac{\partial}{\partial E}
\left( -m_{\star}D_E g_{\star} - 
D_{EE}\frac{\partial g_{\star}}{\partial E} \right),
\label{eqn:stars}
\end{equation}

\begin{equation}
4\pi^2 p(E) \frac{\partial g_{\chi}}{\partial t}=\frac{\partial}{\partial E}
\left( -D_{EE}\frac{\partial g_{\chi}}{\partial E} \right),
\label{eqn:DM}
\end{equation}
with 
\begin{equation}
g_i(E,t)=\int_0^{\infty} f_i(E,t,m)m \mbox{ }dm,
\end{equation}

\begin{equation}
h_i(E,t)=\int_0^{\infty} f_i(E,t,m)m^2 \mbox{ }dm,
\end{equation}
with $ i \in [\star,\chi] $, and the diffusion coefficents can be written as
\begin{equation}
D_{EE}(E)=64\pi^4 G^2 m_{\star} \ln\Lambda \left[ q(E) \int_{-\infty}^{E} 
dE^{\prime}g_{\star}(E^{\prime}) + \int_{E}^{0} dE^{\prime} q(E^{\prime}) 
g_{\star}(E^{\prime}) \right],
\end{equation}
\begin{equation}
D_E(E)=-64\pi^4 G^2 \ln\Lambda \int_{E}^{0} dE^{\prime} 
p(E^{\prime}) g_{\star}(E^{\prime}).
\end{equation}

$ f_{\star} $ is the stellar distribution function, whose evolution 
(Eq. \ref{eqn:stars}) is governed by star-star interactions, and $ f_{\chi} $
is the distribution function for DM and in Eq. \ref{eqn:DM} only DM-star
encounters are considered due to the collisionless nature of DM.

The final steady-state solutions will have the usual $ -7/4 $ slope for stars
(in fact Eq. \ref{eqn:stars} is not different from Eq. \ref{eqn:Fokker_Planck})
and a milder $ -3/2 $ slope for DM, that will be established in the same 
timescale $ t_{rel} $ (Eq. \ref{eqn:relaxation}).
The steeping of the initial profile, due to the presence of a BH, leads to an
increase of the DM annihilation rate. This is why we refer to BHs as
{\it DM Annihilation Boosters}. Although in this case the effect is not
dramatic, since a $ r^{-1.5} $ profile can be found even for models without
BHs (see Sec. \ref{sec:two}), and it is more likely that the overdensity will 
be reduced in a couple of relaxation times \cite{Merritt_Harfst_Bertone} 
(see Sec. \ref{sec:CREST}), we will see cases in the next Sections where BHs
can provide huge boost factors.

The presence of a stellar cusp has been experimentally confirmed for the 
MW \cite{Genzel_et_al,Schodel_et_al}, through the detection of a 
profile with a slope equal to $ -1.4 < \gamma < -1.3 $ in the inner region 
($ r \lesssim 0.38 \mbox{ pc} $) and to $ -2 $ (isothermal profile) in the 
outer region (see Fig. 3). Our Galaxy is a collisional nucleus, since the 
relaxation time is shorter than the age of the Universe 
($ 3.5 \times 10^9 \mbox{ yrs} $ at $ \thickapprox 0.1 \mbox{ } r_h $), so it 
was suggested to interpret its cusp as the Bahcall-Wolf solution to the 
presence of a SMBH with a mass $ \thickapprox 3.7 \times 10^6 M_{\odot} $ 
($ r_h $ is $ \thickapprox 3 $ pc so that the cusp starts more or less where 
$ r=0.1 \mbox{ } r_h $), hypothesis supported also by the luminosity of the 
X-ray source Sgr A$ ^{\ast} $. Anyway, it is more likely that the MW 
experienced a merger between a redshift $ z=2 $ 
\cite{Merritt_Milosavljevic_Verde_Jimenez} and today, so the cusp will
be the result of an ``overdensity regeneration'' (see Sec. \ref{sec:CREST}). 
The actual, detected profile \cite{Merritt_Szell} is consistent with a cusp 
regenerated after a merger occured at a time $ \gtrsim 8 $ Gyr in the past.

The detected inner slope of $ -1.4 $ is not exactly what the Bahcall-Wolf 
solution predicts ($ \gamma=-1.75 $). However, the two results are considered
consistent with each other since the steeper value is derived under the 
simplyfing assumption of a population of stars with identical mass, and if 
the more realistic multi-mass formalism is introduced, the slope will become 
shallower, moving towards the $ -1.4 $ value. The same can be said if, as it 
was argued \cite{Merritt_Szell}, the time required to reach a steady-state 
solution at the Galactic center is $ \gtrsim 10^{10} \mbox{yr} $.

\begin{figure}[tbh]
\begin{center}
\includegraphics[width=0.6\textwidth]{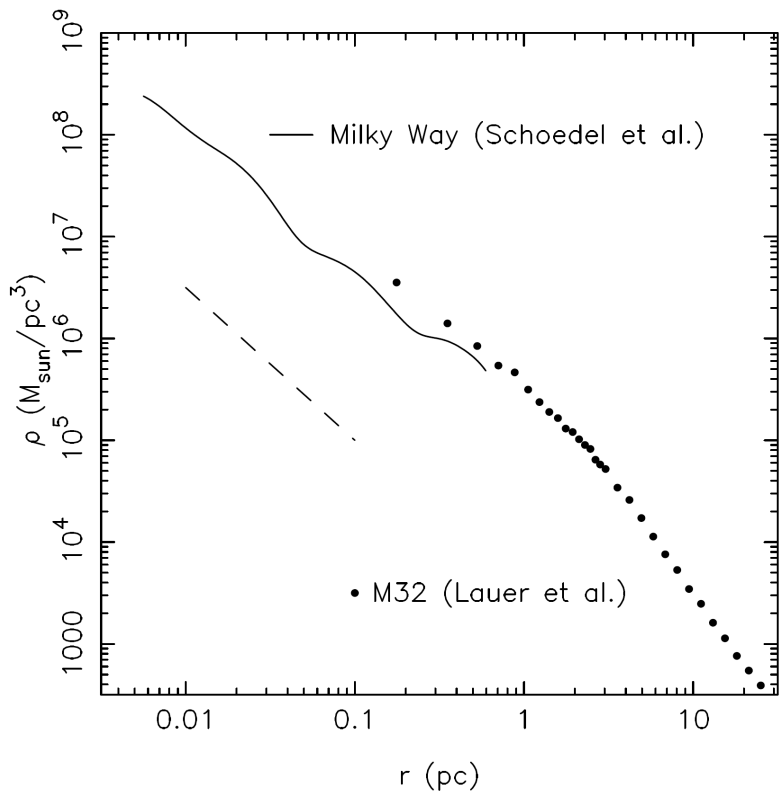}
\label{fig:Milky_Way}
\caption{Mass density profiles near the center of the MW and M32. Dashed line is $ \rho(r) \propto r^{-1.5} $. Both galaxies contain SMBHs with masses $ \sim 3 \cdot 10^6 M_{\odot} $ and with influence radii $ r_h \thickapprox 3 \mbox{ pc} $. Figure taken from Merritt (2006).}
\end{center}
\end{figure}

\subsection{Loss-cone dynamics and BH binaries}
\label{sec:LC_and_binary}

In this Section, we focus on two mechanisms that can reduce, and even 
completely destroy, collisional cusps. First of all, we review the 
basics of {\it loss-cone} dynamics, i.e. the effects related to the presence 
of a BH tidal radius or of the relativistic event horizon. This is only 
partially relevant for DM since it will affect mainly the stellar population, 
but it is relevant in this contest since we have just seen 
(Eq. \ref{eqn:DM}) how stars and DM evolve together, so that a decrease in the 
stellar cusp due to loss-cone \cite{Merritt} will influence the DM density, 
too. Moreover we want to emphasize that any enhancement in the density of DM 
and stars has to survive to a collection of dynamical mechanisms that can 
damp them.

After discussing the loss-cone, we will focus on the possibility that, 
during the merging evolution of a nucleus, a BH binary forms. In this case, 
both stellar and DM distributions will be directly affected 
\cite{Merritt_binaries}, leading not only to the reduction of cusps but also 
to the possible destruction of spikes in models with adiabatic growth 
\cite{Merritt_Milosavljevic_Verde_Jimenez} (see 
Sec. \ref{sec:spike_destruction}).

The loss-cone of a BH is the set of eccentric orbits populated by stars that
are going to intersect the tidal radius.
Such orbits are depleted in a crossing time $ t_{cr} $, since stars are 
eaten up by the BH: tidal forces inside the tidal radius tear a star apart
and these events are accompanied by flares, emissions of light peaked in the
X or UV band with a luminosity of $ \thickapprox 10^{44} \mbox{erg s}^{-1} $.
A handful of these flares have been observed: they have the expected signature
and the number of detections is roughly consistent with theoretical estimates
of the consumption rate  $ \dot{N} $ \cite{Komossa,Komossa_et_al,Halpern_et_al,Wang_Merritt}.

Once depleted, loss-cone orbits can continue to contribute to the consumption
rate only if they are somehow refilled with particles. Energy diffusion
provides a mechanism for such a repopulation. In fact, $ N $-body simulations
confirmed \cite{Merritt} that the zero-flux Bahcall-Wolf solution is
established only approximately and that (for stars) there is a residual flux 
$ F(E) \propto r_t^3/t_{rel}(r_t) $. 

This value is too low compared to the expected $ \dot{N} $ whose main 
contribution comes, instead, from angular momentum diffusion: the 
``classic loss-cone theory'' \cite{Frank_Rees,Bahcall_Wolf,Cohn_Kulsrud} applies to globular clusters (populated with a central BH),
whose relaxation time is so low that they are well-relaxed and old objects. 
The distribution of stars near the BH is therefore assumed to be on a 
steady-state, for which a Fokker-Planck-like formalism is appropriate.
Resulting estimates for the consumption flux can be introduced in the original 
Fokker-Planck equation (Eq. \ref{eqn:Fokker_Planck}) to study how particle 
density is affected by the presence of a tidal radius:
\begin{equation}
4\pi^2 p(E) \frac{\partial f}{\partial t}=-\frac{\partial F_E}{\partial E}
+\rho_{lc}(E,t).
\label{eqn:loss_cone_energy}
\end{equation}

When extended to the study of galactic nuclei, the steady-state approximation
may fail, at least for collisionless nuclei, and the stellar profile
near the tidal radius is, in general, different from the Bahcall-Wolf one.
For example, galaxies are only approximately spherical: their shape is 
more likely to be triaxial and there is the possibility that they are 
governed by centrophillic orbits, i.e. orbits that pass arbitrarily close to 
the BH. In the case that these chaotic orbits survive until late stages in the 
galactic evolution, $ \dot{N} $ would increase, since more particles would
fall into the tidal sphere.

Alternatively, the present galaxy can be the result of cumulative mergers of 
less massive mini-galaxies, each of them hosting a mini-BH: the formation of 
a BH binary would decrease the consumption rate, since all stars with angular 
momentum $ L \lesssim L_{bin}=(2GM_{12}a_h) $ would be ejected 
\cite{Merritt_Wang} ($ M_{12} $ is the total mass of the binary and $ a_h $ 
the major semi-axis when the system becomes ``hard''), preventing loss-cone 
repopulation and leading to lower rates $ \dot{N} $.

Finally, in real galaxies, the diffuse mechanism of orbits refillment 
will cause the nucleus to expand \cite{Merritt} 
\cite{Freitag_Amaro-Seoane_Kalogera,Murphy_Cohn_Durisen,Baumgardt_Makino_Ebisuzaki}, since the density is reduced when 
particles are eaten up and those particles which fall into the loss-cone 
transfer energy to the remaining nucleus with the same effect of a heating 
process. The expansion is visible in one single relaxation time, the 
``decay'' goes on at a constant velocity and the density can be written as 
$ \rho(r,t)=\rho_c(t)\rho^{\ast}(r) $, where $ \rho^{\ast}(r) $ is the initial 
profile, while $ \rho_c(t) \propto t^{-1} $. As a consequence, present-day
collisional nuclei could have been denser in the past.

It has been suggested that also the presence of a BH binary can effectively 
reduce the Bahcall-Wolf cusp. The growth of a galaxy is thought to pass 
through the agglomeration of smaller galaxies and protogalactic fragments. If 
more than one of these subhalos contain a BH, the two objects will form a 
binary system whose dynamics can strongly affect stars and DM. This scenario 
has received great attention since mergers and the ultimate coalescence of the 
BH binary are ideal targets for the detection of gravitational waves 
\cite{Thorne_Braginskii}. Evidences for the presence of such binaries can be 
found in Ref. \cite{Komossa_et_al_GW} and are based on the detection of 
multiple active nuclei in the same galaxy \cite{Rodriguez_et_al}.

Consider a compact object with mass $ M_2 $ moving, with its nucleus, around a 
BH with mass $ M_1 $, being $ q=M_2/M_1<1 $ the mass ratio and $ M_{12} $ the 
total mass.
The evolution of the binary can be described by three different phases
\cite{Merritt_binaries,Merritt}: first, the smaller BH decays due to 
the dynamical friction with stars of the other nucleus, and the separation 
$ R_{12} $ between the two objects drops down. When the influence radius 
$ r_h $ of the more massive BH is reached, the two objects can be considered 
as a bound object and the first phase comes to an end. The infall time scale 
\cite{Merritt} suggests that binaries are not so uncommon since, for 
reasonable values of $ q $ ($ q \thickapprox 10^{-3} $ and 
$ M_{1} \thickapprox 10^8 M_{\odot} $), they are able to form before the
Hubble time.

The second phase is characterized by a quick ``shrinking'' of the binary, 
until it becomes {\it hard}, i.e. the binding energy equals the kinetic 
energy, or equivalently the major semi-axis reaches \cite{Merritt_binaries}
$ a_h=q/(1+q)^2 \cdot r_h/4 $. The third phase, when $ a \lesssim a_h $, is
the least known: a binary in a fixed background begins to harden at a constant 
hardening rate $ s=d(1/a)/dt $, but physical binaries has already ejected 
almost all stars on intersecting orbits and the rate suddendly drops. These 
orbits need to be repopulated, usually by energy diffusion, but this effect 
is more likely to be only subdominant, at least in those bright galaxies where 
the scouring of BH binaries has been detected, characterized by a relaxation 
time higher than the Hubble time.
In numerical simulations with finite $ N $, gravitational encounters will 
unphysically continue to supply particles to the binary at rates roughly 
proportional to $ N $, so experimentally it is more usefull to define the 
semi-major axis $ a_{stall} $ where the hardening rate goes to zero 
\cite{Merritt_binaries}.
From $ N $-body simulations it results
\begin{equation}
\frac{a_{stall}}{r_h^{\prime}}=0.2 \frac{q}{(1+q)^2},
\end{equation}

$ r^{\prime}_h $ is a second influence radius, defined as the radius where
the total mass of particles within $ r^{\prime}_h $ after the binary has 
stalled is equal to twice $ M_{12} $.
This values for $ a_{stall} $ is a couple of orders of magnitude higher than
the distance where the binary coalescences.
This is known as the ``final parsec problem'', since evidence is strong 
that BHs binaries do efficiently coalescence \cite{Rodriguez_et_al,Merritt_Ferrarese} at last. Many solutions have been proposed
\cite{Gualandris_Merritt_one}, e.g. that, as for loss-cone orbits, the 
presence of centrophillic orbits in realistic triaxial galaxies can affect 
the above considerations, so that the binary continues to shrink even to 
$ a \lesssim a_h $. 

Assuming that the binary does stall at $ a_{stall} $, it will have transferred
an energy
\begin{equation}
\Delta E \thickapprox -\frac{GM_1M_2}{2r_h}+\frac{GM_1M_2}{2a_h} \thickapprox
-\frac{1}{2}M_2 \sigma^2 + 2 (M_1+M_2)\sigma^2 \thickapprox 2(M_1+M_2)\sigma^2,
\end{equation}
to the particles in the nucleus. This relation has been used to explain
the mass deficit in the core of brightest galaxies since such an energy
release will let stars leave the central core, with a total displacement
of mass $ M_{def} $ and $ 0.4 \lesssim M_{def}/M_{12} \lesssim 0.6 $ 
\cite{Merritt_binaries} for $ 0.05 \lesssim q \lesssim 0.5 $. 

Strictly speaking, the observed mass deficits $ M_{def} $ reach values that 
are even four times larger than the mass of the binary $ M_{12} $ (that, if 
the coalescence occurs, is also the mass of the final BH). We can account for 
values as large as $ M_{def}/M_{12} \lesssim 2 $ if the nucleus experiences 
more than one merger, with more than one binary forming. The total mass 
displaced will be simply the sum of each $ M_{def} $ during each single 
merger. For even larger values, other mechanisms have to be evoked, e.g. 
the possibility that a third BH arrives when the first two have not 
coalescenced yet. In such a situation, one of the BH usually leaves the 
nucleus ({\it gravitational slingshot} effect) leading to higher values for 
$ M_{def} $. Similarly one of the SMBH of a BH binary can be expelled with a 
high velocity, due to the so-called {\it gravitational-wave rocket} effect 
\cite{Gualandris_Merritt_two}.

\section{Adiabatic growth of Black Holes}
\label{sec:four}

\subsection{Adiabatic growth of Black Holes}
\label{sec:adiabatic_growth}
In this Section we relax the assumption of time-independent gravitational 
potential. As we will see, this can, in some cases, lead to large DM
overdensities. In particular, the adiabatic growth of BHs can produce the
steepest DM profiles discussed in literature. 

The seed BH grows in an already-formed nucleus with a stable (stellar 
or DM) configuration. The condition of adiabaticity guarantees that the growth 
timescale is large compared to the crossing time, but smaller than the 
relaxation time (for the stellar distribution). As a consequence, nuclei 
where a BH have grown adiabatically have not yet reached a stable, relaxed 
stellar configuration. 
Another consequence \cite{Binney_Tremaine} is that the angular momentum 
and the radial action (i.e. $ J_r=\oint v_r dr $, where $ v_r $ is the 
radial velocity and the integral is over one closed orbit) are conserved. 
The hierarchy between the BH accretion timescale and the nuclear crossing time,
that lies at the core of the adiabatic assumption, is reasonable at least for
BHs with masses $ M_{\bullet} \lesssim 10^{10} M_{\odot} $, as can be checked
adopting the shortest timescale for the BH growth, i.e. the Salpeter time 
$ t_s=M_{\bullet}/\dot{M}_{Edd} $ (where $ \dot{M}_{Edd} $ is the
Eddington accretion rate), and comparing it with the crossing time at the
influence radius $ t_{cr} \propto M_{\bullet}\sigma/r_h $, using the 
$ M_{\bullet}-\sigma $ relation (See Eq. \ref{eqn:M_sigma}).

The first study on the impact of adiabatic growth \cite{Young} on stars, 
analyzed the case of a 
non-singular isothermal stellar  profile, and predicted an overdensity 
extending to the same size of the initial core, with a slope equal to 
$ -3/2 $. In the case of a DM halo, such overdensity has been called 
{\it spike} \cite{Gondolo_Silk}, to distinguish it from the aforementioned 
DM cusps.

A numerical algorithm that mimic adiabatic growth was also developed, in order
to confirm the creation of the overdensity 
\cite{Quinlan_Hernquist_Sigurdsson}. The method is very flexible and, in fact, 
it was applied to initial models other than the isothermal distribution 
\cite{Quinlan_Hernquist_Sigurdsson}. Two classes can be identified: the 
first includes all those profiles called ``analytic cores'', characterized by 
a density that can be expanded in a power law series near the BH 
($ \rho(r) \thickapprox \rho_0+1/2 \rho^{\prime\prime}_0 r^2+\dots$), while
the second describes the so-called $ \gamma $ models that exhibit a 
power law density profile in the inner region: $ \rho(r) \propto r^{-\gamma} $.

As benchmark cases, the $ \gamma $-models with $ 0 \leq \gamma \leq 2 $ 
and the isothermal model (as an example of analytic profiles) are considered
here, and results are presented in Tab. I and in Fig. 4 
\cite{MacMillan_Henriksen} (only for the isothermal and for $ \gamma=1 $).

The spike radius $ r_{sp} $, i.e. the  distance where the slope changes due to
the presence of the BH, depends on the BH mass and it is related to the 
influence radius as $ r_{sp} \thickapprox 0.2 \mbox{ } r_h $ 
\cite{Merritt_Carnegie}.
Inside such radius, the spike has a slope $ \gamma_{sp} $, that 
depends on the initial $ \gamma $. In the case of a model with analytic core 
the final slope is $ -3/2 $ \cite{Young}, while for the $ \gamma $ models an 
analytic relation holds \cite{Peebles}
\cite{Quinlan_Hernquist_Sigurdsson,Ullio_Zhao_Kamionkowski}:
\begin{equation}
\gamma_{sp}=\frac{9-2\gamma}{4-\gamma}.
\label{eqn:gamma_spike}
\end{equation}

Such relation is valid under the following assumptions (satisfied by all the 
models in Quinlan {\it et al.} \cite{Quinlan_Hernquist_Sigurdsson}): 
\begin{itemize}
\item the distribution function is isotropic; 
\item the gravitational potential can be written as $ r^{2-\gamma} $ in the 
$ r \rightarrow 0 $ limit;
\item $ f $ diverges as $ \left[ E-\Phi(0) \right]^{-n} $ in the limit 
$ E \rightarrow \Phi(0) $ (this last requirement is what makes a model
with analytic core different from a $ \gamma $ model). 
\end{itemize}

\begin{table}[thb]
\begin{center}
\begin{tabular}{|c|cccc|}
\hline
Model & $ \gamma $ & $ n $ & $ \gamma_{sp} $ & $ C $ \\
\hline
isothermal & 0 & 0 & 3/2 & 9/4 \\
$ \gamma $ model ($ \gamma = 0 $) & 0 & 1 & 2 & 9/4 \\
$ \gamma $ model ($ \gamma = 1 $) & 1 & 5/2 & 7/3 & 7/3 \\
$ \gamma $ model ($ \gamma = 3/2 $) & 3/2 & 9/2 & 12/5 & 12/5 \\
$ \gamma $ model ($ \gamma = 2 $) & 2 &  & 5/2 & 5/2 \\
\hline
\end{tabular}
\end{center}
\label{tab:slopes}
\caption{Different quantities computed from the adiabatic growth of the initial models proposed by Quinlan, Sigurdsson and Hernquist (1995). $ \gamma $ and $ \gamma_{sp} $ are the initial and final slope in the density profile for the region closer to the BH. $ n $ indicates how the distribution function diverges as $ E \rightarrow \Phi(0) $ and $ C $ is the slope of the final density profile if it were made of particles on circular orbits. The value for $ n $ in the $ \gamma=2 $-model is absent since the equation used to derive $ n $ is not valid for $ \gamma=2 $, but for $ \gamma \rightarrow 2 $ the final profile has $ \gamma_{sp} \rightarrow 5/2 $.}
\end{table}

\begin{figure}[tbh]
\begin{center}
\includegraphics[width=6.5cm]{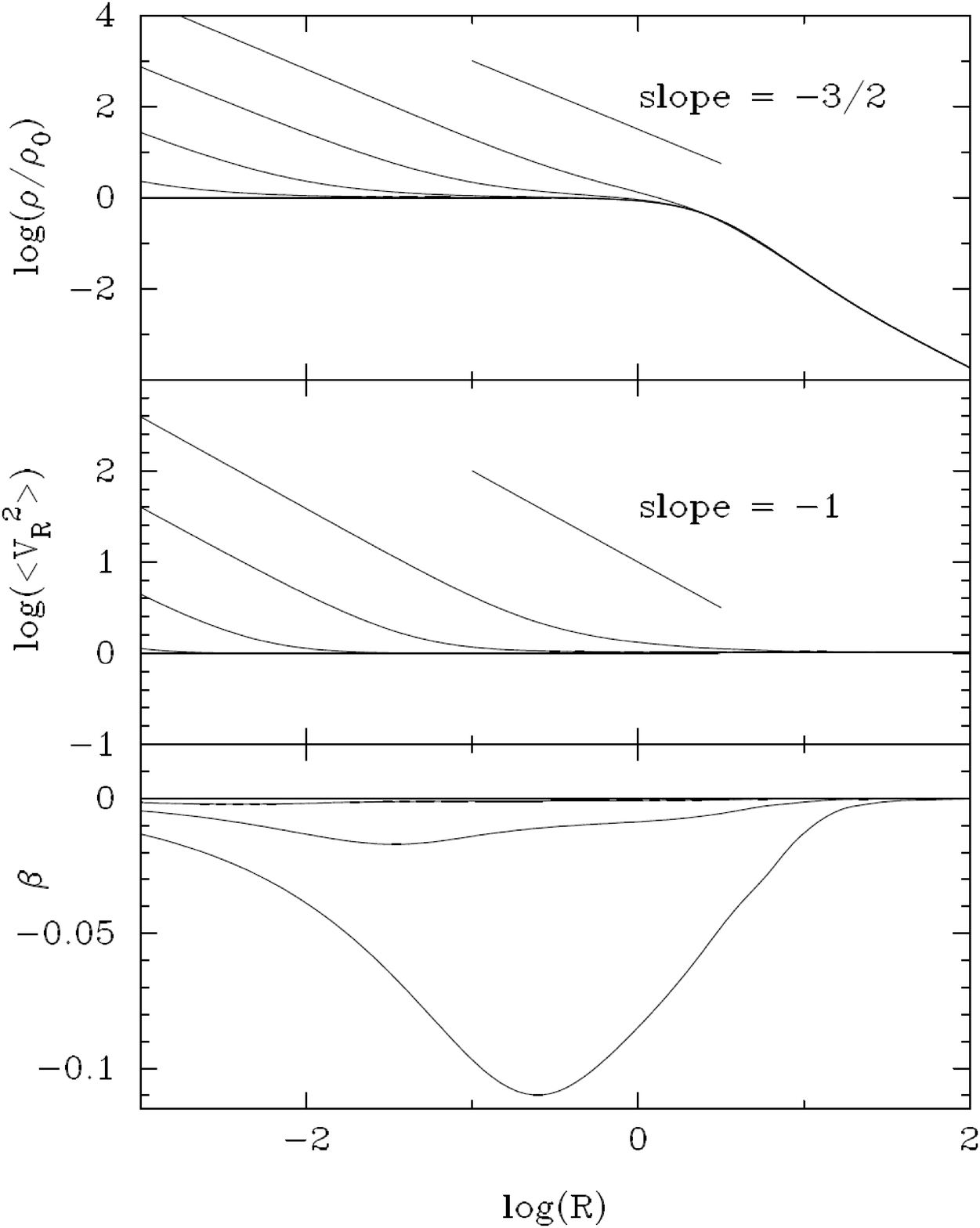}
\includegraphics[width=6.5cm]{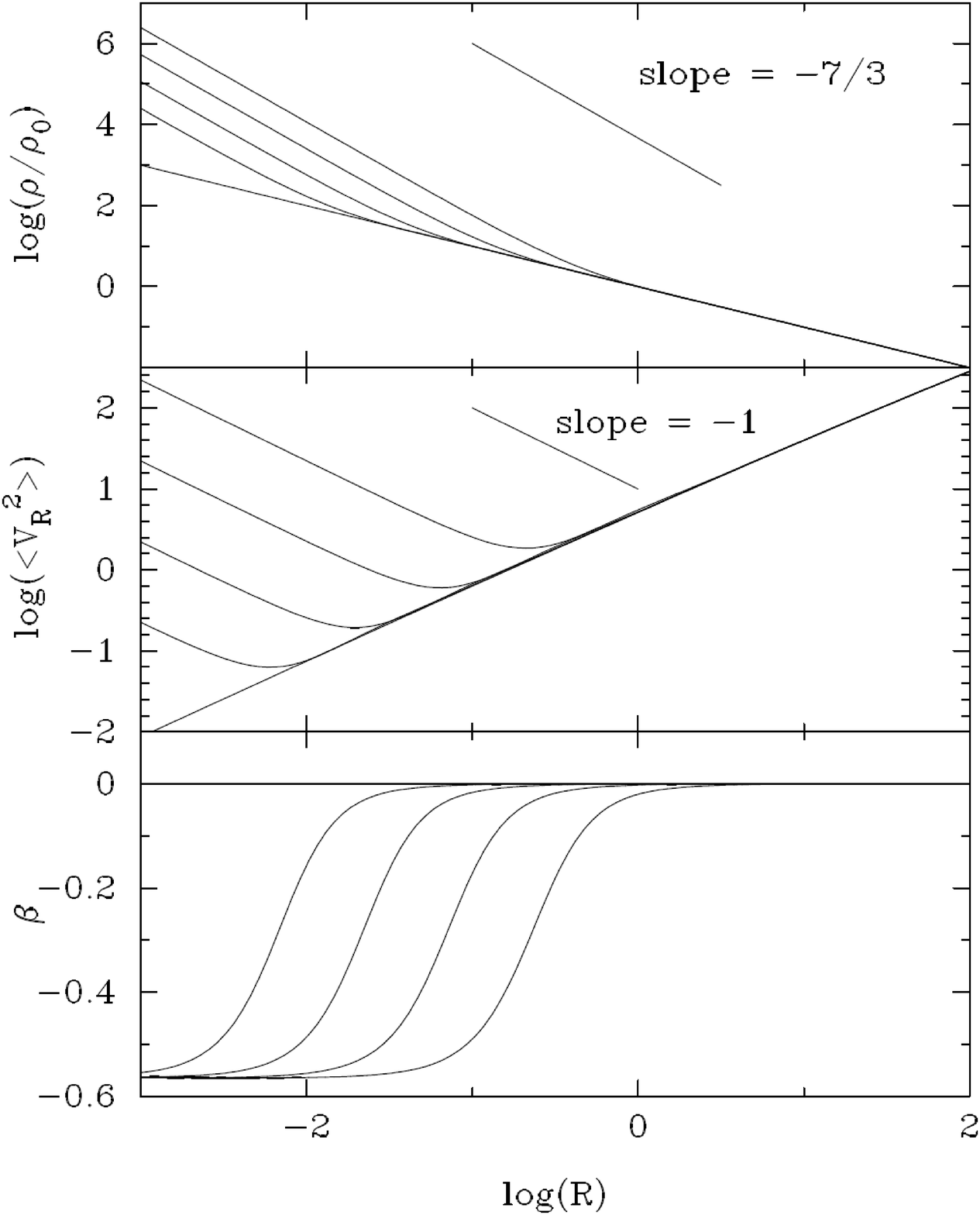}
\label{fig:adiabatic_growth}
\caption{Adiabatic growth in the isothermal ({\it left panel}) and for a $ \gamma = 1 $ ({\it right panel}) model for a value of $ M_{\bullet} $ of 0.001, 0.01, 0.1 and 1.0 (mass units explained in the Reference), with the mass encreasing from bottom to top in the top two panels and from top to bottom in the last panel. The density is shown in the top frame, the averaged radial velocity in the middle and the anisotropy parameter $ \beta $ in the last frame. $ \beta=1-\langle v_t^2 \rangle / \langle 2 v_r^2 \rangle $ where $ v_t $ ($ v_r $) is the tangential (radial) component of velocity. Figure taken from MacMillan, Henriksen (2002).}
\end{center}
\end{figure}

Comparing the first two lines in Tab. I, it can be seen that, even if both 
models start with a costant core, they develop very different final spikes, 
due to the different behaviour of the distribution functions in the 
$ E \rightarrow \Phi(0) $ limit, suggesting that the formation of a 
strong spike is not a consequence of a singularity in the density profile 
but in the distribution function, and in particular in the way cold orbits 
(populated by stars with a low velocity) are arranged (see Sec.
\ref{sec:spike_destruction}).
We will not consider initial configurations with $ \gamma $ larger than 1.5 
(see Sec. \ref{sec:two}), so the steepest spike has a slope of $ -12/5 $ 
(when $ \gamma=1.5 $). 

The velocity dispersion reacts to the BH growth in a similar way for both
classes of initial models: in fact, a Keplerian rise appears, with a slope 
of $ -1/2 $. On the contrary, the anisotropy is substantially different: 
analytic models exhibit a mild tangential anistropy at an intermediate 
distance from the BH but remains isotropic at the center; the more massive the 
BH is, the higher the anistropy. While for $ \gamma $ models, orbits are 
tangentially-biased in the central region and the anisotropic area increases 
with more massive BHs. If the hypothesis of an isotropic distributions is 
relaxed, a nucleus made interely by circular orbits evolves to a profile with
final slope equal to $ C $, shown in Tab. I. As one can see, for $ \gamma $ 
models $ \gamma_{sp}=C $, while the circular isothermal model exhibits a much 
steeper slope, although with a value not higher than for the $ \gamma $ 
models. Such consideration suggests that results from adiabatic growth are not 
very sensible to possible violations of isotropy in the initial configuration.

Even if spikes are the steepest known overdensities, they cannot be 
considered as signatures of BH growing adiabatically (since a simple singular 
isothermal profile that nothing has to do with adiabatic growth is steeper 
than half of the models in Tab. I) and neither hints of the presence of a 
central BH (since a $ -1.5 $ slope, as in the first line of Tab. I, can be 
found for halos without BH). On the contrary, a rise in the $ \sigma $ profile 
is quite a robust indication of the presence of a central object.

\subsection{Destruction of spikes}
\label{sec:spike_destruction}
The formation of spikes described in the previous Section leads to the largest
annihilation boost factors: the mechanism of adiabatic growth can, in fact, 
produce inner slopes as steep as $ -2.25 $ (See Tab. I), for profiles that 
will be characterized by a large annihilation flux (which is proportional to 
the integral of the DM density squared), with interesting consequences for 
Indirect DM searches.

However, it has been argued \cite{Ullio_Zhao_Kamionkowski} 
\cite{Bertone_Merritt,Merritt_Milosavljevic_Verde_Jimenez}, that the
formation of spikes requires {\it ad hoc} initial conditions for
the DM halo. Moreover, even if spikes do form, then, dynamical effects can 
reduce or even destroy them, as we considered before for collisional cusps.
A spike can form even from an initial density profile that does not diverge
\cite{Ullio_Zhao_Kamionkowski}, but in order to produce a significant 
overdensity, the distribution function of cold orbits has to diverge in the
$ E \rightarrow \Phi(0) $ limit. In fact, cold orbits are those which provide 
the particles that will form the spike, since they are the most affected by 
the presence of the central BH.

But these cold orbits are more likely to be depleted due to the interactions of
stars with molecular clouds or globular clusters or other bodies that 
can pass through a galactic nucleus. Moreover, the evolution of a galactic 
nucleus is thought to be characterized by the cumulative mergers of 
sub-nuclei and even a single merger event can have dramatic consequences 
on the distribution of cold orbits. In other words, one can enumerate a 
collection of effects that effectively heat up the particles near the BH, 
so that thay can leave the central region obstacling the formation of the 
spike.

Also in the case that the spike is formed, it is unlikely that it will survive 
to the evolution of the nucleus and, in particular to the presence of 
dynamical mechanisms that would provide an additional heating source to 
particles on cold orbits, with the result of highly reduce of even destroy the 
enhancement.
Numerical simulations have been performed in order to quantify these effects:
for example the possibility that the BH forms slightly off the center of the
nucleus was described by Ullio, Zhao and Kamionkowski 
\cite{Ullio_Zhao_Kamionkowski}. The BH would slowly spiral in, towards the 
center \cite{Nakano_Makino}, and then adiabatically grow to the final value. 
But if the initial value for the BH mass is too low, the spiraling would take 
too long to finally reach the center, while, if the BH is too massive, its 
scouring effect on the DM particles would flatten the central density, to 
values that can be even lower than the initial profile (see 
Fig. 5). 

Moreover, gravitational interactions of DM particles with baryons in stars
modify the evolution of DM in the spike, reducing the enhancement, in the 
same way that stars heat the DM particles in a collisional cusp causing 
its damping (see Sec. \ref{sec:CREST}) \cite{Merritt_Harfst_Bertone} 
\cite{Ullio_Zhao_Kamionkowski}.
Simulations on the effects of galactic mergers can also be found in
Ref. \cite{Merritt_Milosavljevic_Verde_Jimenez}. 
Other objections have been put forward, suggesting that spikes can form only
as results of a series of accidents and, therefore, are not expected to be 
common in the local Universe \cite{Ullio_Zhao_Kamionkowski}.

\begin{figure}[tbh]
\begin{center}
\includegraphics[width=8cm]{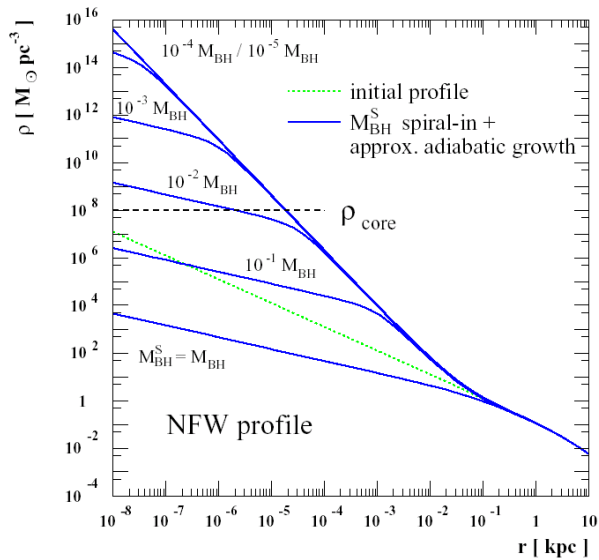}
\label{fig:off_center_formation}
\caption{Modification of a NFW DM density profile due to the off-center formation of a BH seed of mass $ M_{BH}^{S} $, its spiraling towards the center of the DM distribution, and its adiabatic growth to the present-day mass of the BH $ M_{BH} $ at the Galactic center. Several different values of the BH seed mass are plotted. $ \rho_{core} $ is the maximum WIMP density above which WIMPs are depleted by pair annihilations. Figure taken from Ullio, Zhao and Kamionkowski (2001).}
\end{center}
\end{figure}

\subsection{Regeneration of cusps}
\label{sec:CREST}
In the previous Sections we focused on processes that can destroy 
overdensities (whether cusps or spikes). However, gravitational interactions 
among particles (during the evolution of the nucleus) can partially regenerate 
such structures. In a realistic nucleus populated by stars 
and DM, star-star collisions and star-DM collisions (given enough time) drive 
the evolution towards a steady-state $ -7/4 $ profile for the baryonic 
component and a $ -3/2 $ profile for the DM component 
\cite{Merritt_Harfst_Bertone} (see Section \ref{sec:Bahcall_Wolf}).
This is true also if the nucleus is the result of an early evolution phase 
in which previous enhancements were destroyed. In other words, consider a 
nucleus with a short $ t_{rel} $ and a steep DM and stellar profile (due to
collisional relaxation or adiabatic growth of the central BH). If a merger 
occurs and a BH binary forms, the displaced mass will reduce or even destroy 
both overdensities.
But, due to the short relaxation time, it can happen that the nucleus has 
enough time to reconstruct, from the core profile after the BHs coalescence, 
the collisional solutions. The new DM cusp is called CREST (Collisionally 
REgenerated STructure). This idea can be checked analitically, applying the
two-body Fokker-Plack formalism to a core profile describing a nucleus after 
the scouring of a BH binary, but also with the help of numerical routines 
\cite{Merritt_Harfst_Bertone}. Results are summarized in  
Fig. 6: DM CRESTs are not as steep as spikes, but they have the advantage to 
form from very general initial profiles, given that the (stellar) relaxation 
time is short enough. They need a timescale of roughly 
$ t_{rel}(0.2 \mbox{ }r_h) $ to form, but then DM particles in the CREST 
continue to be heated by gravitational interactions with stars and the 
$ -3/2 $ solution, therefore, decays in a self-similar way
\begin{equation}
\rho_{\chi}(r,t) \thickapprox \rho_{\chi,0}(r)G(t/t_{rel}),
\end{equation}
with $ dG/dt<0 $, so that after  $ 4.5 \mbox{ } t_{rel}(r_h) $ the reduction 
is of a factor $ 1/e^2 $.

The balance between the requirement that the relaxation time is short enough
to let the CREST form but not too short to make the CREST not to decay too 
much, leaves us with a rather narrow window of galaxies where CRESTs can be 
present: one can detect them in galaxies with a luminosity
$ 3 \cdot 10^8 L_{\odot} \lesssim L \lesssim 3 \cdot 10^9 L_{\odot} $.
The MW is inside this range and, in fact, many proposed to interpret the cusp 
detected for our Galaxy \cite{Schodel_et_al} as a reconstructed structure 
after a merger occured $ \sim $ 8 Myr ago \cite{Merritt_Szell}.

\begin{figure}
\begin{center}
\includegraphics[width=12cm]{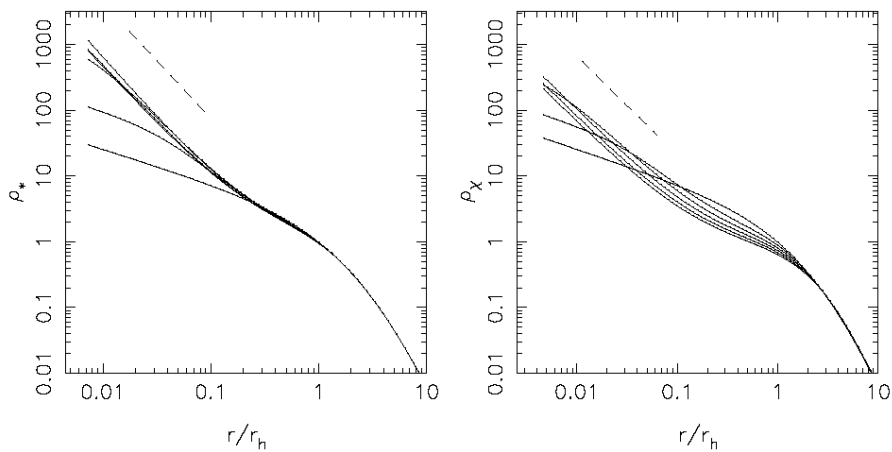}
\label{fig:FP_crest}
\caption{Solutions of the Fokker-Planck equation that describe the joint evolution of stars and dark matter around a BH due to star-star and star-DM gravitational encounters. Length unit is $ r_s $, while the density is in units of the initial value at $ r_h $. Curves show the stellar (left) and DM (right) density profiles at time 0, 0.2, 0.4, 0.6, 0.8, 1.0 in units of the initial relaxation time at $ r_h $. Dashed lines are the steady-state solutions. Figure taken from Merritt, Harfst and Bertone (2006).}
\end{center}
\end{figure}

\section{Intermediate Mass Black Holes}
\label{sec:five}
\subsection{Formation scenarios}
\label{sec:formation}
The steep DM slopes produced by the adiabatic growth of BHs make these objects
extremely interesting since they may effectively act as DM Annihilation 
Boosters. This circumstance encouraged many authors to look for possible 
ways to evade the dynamical effects causing spikes to damp.

Two possibilities have been proposed: the first is to focus on IMBHs instead 
than of SMBHs. IMBHs can be present within substructures of DM halos 
\cite{Koushiappas_Bullock_Dekel,Miller_Colbert} and their 
evolution is such that the objections raised for spikes around SMBHs
do not apply \cite{Bertone_Zentner_Silk}.
The second is considering the contribution of spikes and mini-spikes (DM
overdensity around IMBHs) to the diffuse Extragalactic Gamma ray Background 
(EGB), integrating the signal from high redshift, i.e. when spikes were 
already formed but destruction mechanisms were not yet effective 
\cite{Ahn_Bertone_Merritt_Zhang,Horiuchi_Ando,Ando}.

We focus on IMBHs first. Their masses range from $ \thickapprox 20 $ to 
$ 10^6 M_{\odot} $ and adiabatic growth leads to the formation of 
``mini-spikes'' in the same way as with SMBHs.
We don't have direct observational evidence for the existence of IMBHs, but 
some hints come, e.g., from Ultra Luminous X-ray sources (ULXs) 
\cite{Swartz_Ghosh_Tennant_Wu}, sources that emit in the X band with a 
luminosity higher than $ 10^{39} \mbox{erg s}^{-1} $, and, hence, not 
compatible with the interpretation as BHs accreting at the Eddington limit.
But, due to their positions in the host galaxy, they cannot be explained
in terms of AGNs either. The hypothesis of a BH with a mass higher than 
$ 15-20 M_{\odot} $ and less massive than a SMBH seems to be a fair 
explanation, instead.

Many authors also proposed that globular clusters can host IMBHs, and a
possible confirmation of this hypothesis comes from the fact that the mass scale
for an IMBH and the value of the stellar velocity dispersion measured in
globular clusters fall exactly at the extrapolation at lower values of
the $ M_{\bullet}-\sigma $ relation valid for SMBHs \cite{Miller_Colbert}.
From a theoretical point of view, IMBHs can also help to 
explain the formation of SMBHs: the Sloan Digital Survey \cite{Fan_et_al} 
\cite{Barth_Martini_Nelson_Ho,Willott} has detected quasars up to  
redshift $ z \thickapprox 6 $ suggesting that SMBHs were already present 
when the Universe was $ \sim $ 1 Gyr old. One of the most natural way to 
understand this is that SMBHs grew, through a phase of fast accretion 
and mergers, starting from already massive seeds. In fact, a generic 
prediction of scenarios that seek to explain the properties of the observed 
SMBH population, is that a large number of ``wandering'' IMBHs exist in DM 
halos \cite{Islam_Taylor_Silk_first,Volonteri_Haardt_Madau} 
\cite{Koushiappas_Bullock_Dekel}.

Despite their theoretical interest, it is difficult to obtain conclusive 
evidence for the existence of IMBHs. A viable detection strategy could be 
the search for gravitational waves produced in the mergers of the IMBHs
population \cite{Thorne_Braginskii,Flanagan_Hughes_first,Matsubayashi_Shinkai_Ebisuzaki,Koushiappas_Zentner}, with 
space-based interferometers such as the Large Interferometric Space Antenna 
LISA \cite{LISA_site}.

Two formation scenarios are discussed here, follwing Ref. \cite{Bertone_Zentner_Silk}.
In the first (scenario A), IMBHs form from the gravitational collapse of 
Population III stars, which are usually heavier than local stars,
since they grow in an environment with very low metallicity, for which metal
line cooling can be neglected. As a consequence, the Jeans mass (that scales
with the temperature as $ T^{3/2} $) is higher, allowing the formation
of more massive structures. Such stars are characterized by very low 
metallicity, too, meaning that they will lose little of their mass due to 
winds and weak pulsations. Population III stars with masses above larger than 
$ 250 M_{\odot} $ would be able to collapse directly to BHs without 
any explosion \cite{Miller_Colbert}.

In the second scenario (scenario B), the formation starts at high redshift 
($ z \thickapprox 15 $) from a gas cloud that is massive enough to cool down 
forming a protogalactic pressure-supported disc at the center of the cloud, 
made mainly by stars that, according to the distribution of angular 
momentum in the cloud, are in the low momentum tail of the distribution.
The dynamics of the disc is governed by an effective viscosity that transfers 
matter at the center and angular momentum outward. This flow will go on
until the first stars start to fragment in the outer region of the disc:
the so-formed central object will undergo gravitational collapse, forming the 
final BH, with a mass scale of $ \thickapprox 10^5 M_{\odot} $ (corresponding 
to an initial cloud with virial mass $ M_{vir} \thickapprox 10^7 M_{\odot} $) 
or \cite{Bertone_Zentner_Silk}

\begin{equation}
M_{\bullet} = 3.8 \times 10^4 M_{\odot} \left( \frac{\kappa}{0.5} \right)
\left( \frac{f}{0.03} \right)^{3/2} \left( \frac{M_{vir}}{10^7 M_{\odot}}
\right) \times \left( \frac{1+z_f}{18} \right)^{3/2} \left(
\frac{t}{\mbox{10 Myr}} \right),
\end{equation}

where $ f $ is the fraction of the total baryonic mass in the gas cloud that 
has cooled into the disc, $ \kappa $ is that fraction of the baryonic mass 
of the disc that forms the final BH, $ M_{vir} $ is the halo virial mass, 
$ z_f $ is the redshift when the formation starts from the cloud and $ t $ the 
timescale for the evolution of the first generation of stars. 

Regardless of the particular formation scenario, a population of IMBHs
is predicted, each of them surrounded by a mini-DM halo and a mini-spike. 
These mini-halos evolve, merge with each other and form the actual big 
halos of galaxies. The hosted IMBHs will merge too, possibly contributing to
the formation of the central SMBH. However a fraction of initial mini-halos 
never experiences any mergers and the associated DM overdensity can survive
till today.

Moreover, at least for scenario B, IMBHs form exactly at the center of the 
baryonic distribution in the mini-halos, forbidding BH off-center formation. 
Strictly speaking, in order to avoid off-center formation, one should require 
that the pristine BH forms at the center of the DM distribution and this does 
not necessarily coincide with the center of baryons, since stars, been 
collisional, can experience a different evolution than the collissionless DM, 
resulting in a net displacement between the two distributions. But mini-halos 
are supposed to have a very low baryonic content, with no violent interactions 
able to drive the two distributions away one from the other: the fact that 
mini-halos are made most entirely by DM solves the possible off-center 
formation and, at the same time, the problem related to stars-DM interactions. 
Scenario B is thus able to circumvent all the mechanisms for spike destruction
described above, while in scenario A they are reduced but still efficient: 
in the following Sections, we are going to consider scenario A as a 
conservative model for IMBHs, instead of the more optimistic scenario B.

For the MW, an IMBHs population of $ 1027 \pm 84 $ objects is predicted
by the numerical simulations in Ref. \cite{Bertone_Zentner_Silk}, with a mass 
of approximatively $ 10^2 M_{\odot} $ (scenario A), while for the scenario B 
the MW hosts just $ 101 \pm 22 $ with a distribution in mass centered on 
$ 10^5 M_{\odot} $ and log-normally distributed ($ \sigma_{\bullet}=0.9 $), 
as it can be seen in the left panel of Fig. 7.

\subsection{Gamma-rays from DM annihilations around IMBHs}
\label{sec:IMBHs}
\begin{figure}
\begin{center}
\includegraphics[width=6.5cm]{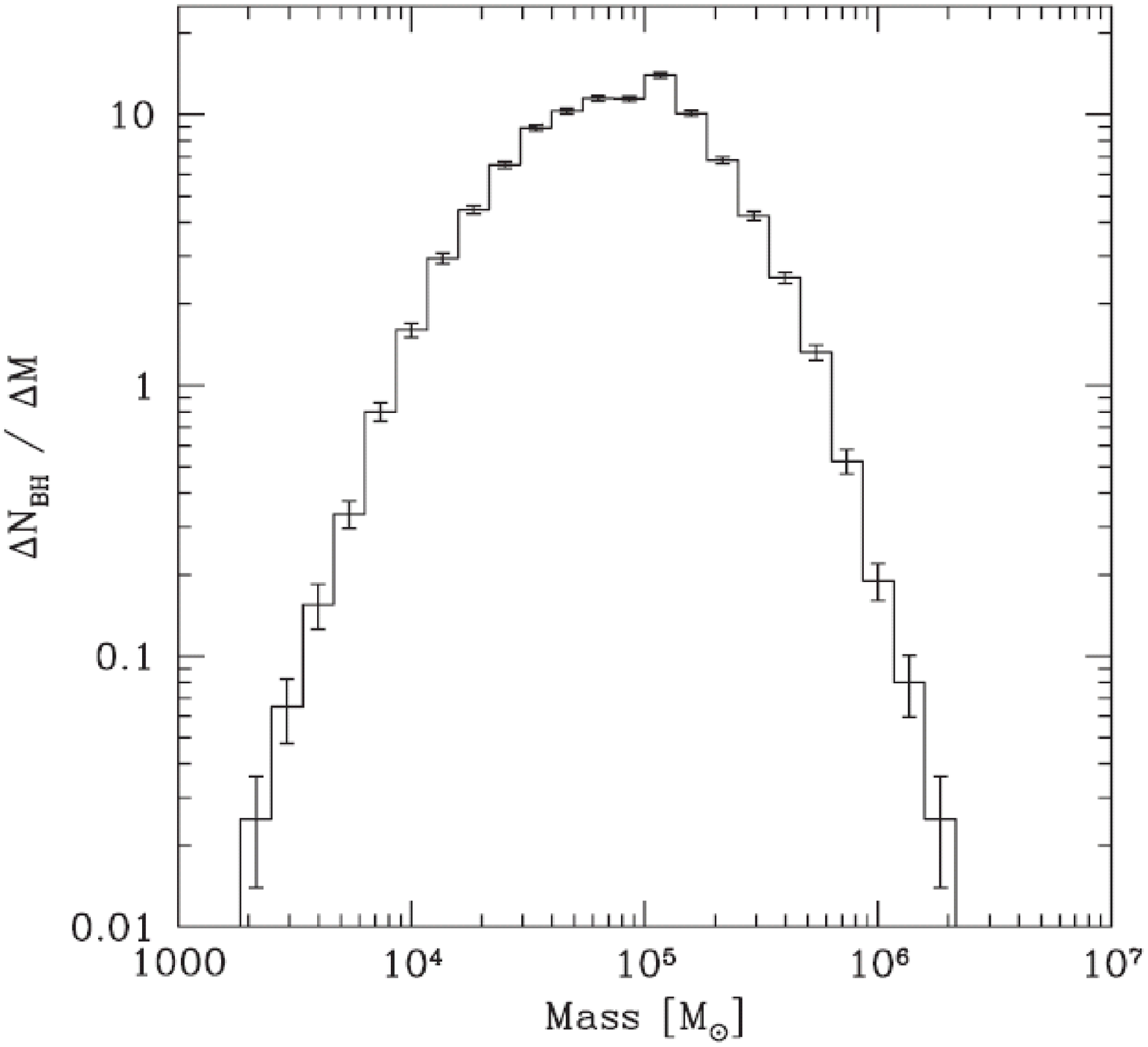}
\includegraphics[width=6.5cm]{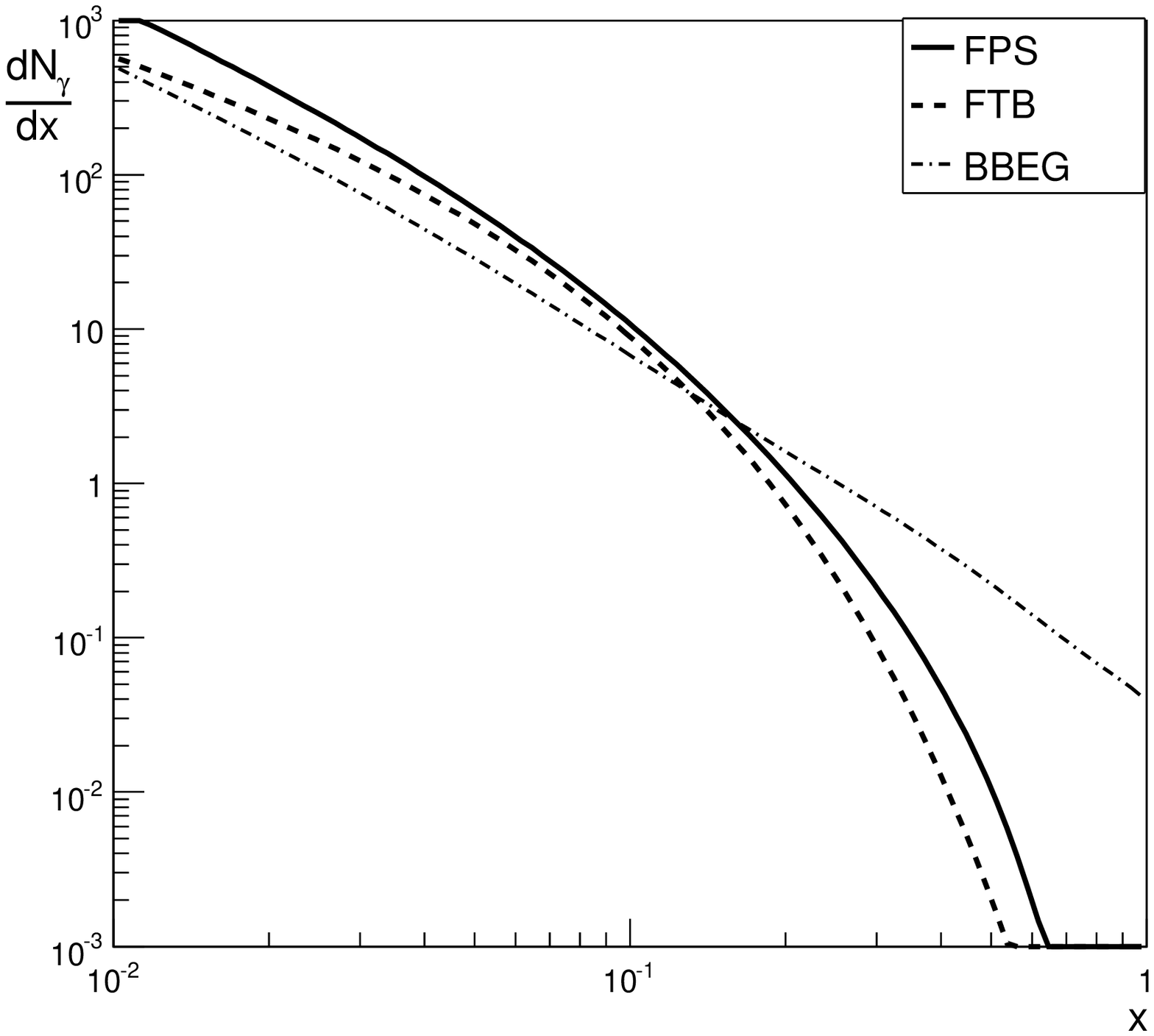}
\label{fig:Bertone_Zentner_Silk}
\caption{{\it Left panel:} Mass function of unmerged IMBHs in scenario B, for a MW halo at $ z=0 $, averaged over 200 Monte Carlo realizations of a MW-like halo with virial mass $ M_{vir}=10^{12.1}h^{-1}M_{\odot} $. Figure taken from Bertone, Zentner and Silk (2005). {\it Right panel:} Differential photon spectrum per annihilation. Different parametrizations and annihilation channels are shown. The solid line is an analytic fit to the $ b \bar{b} $ channel, as obtained by Fornengo, Pieri and Scopel (2004) with $ m_{\chi}=1 \mbox{ TeV} $. The dashed line is relative to the same annihilation channel, but with a different parametrization of the Fragmentation Functions obtained by Fornasa, Taoso and Bertone (2007). The dotted line is relative to UED DM and includes FSR from annihilation to charged leptons. Figure taken from Fornasa, Taoso and Bertone (2007).}
\end{center}
\end{figure}

The prospects for indirect detection of DM from mini-spikes around IMBHs populating the MW
halo has been studied by Bertone, Zentner and Silk 
\cite{Bertone_Zentner_Silk}. They considered 200 different statistical 
realizations of the MW, obtaining an IMBH catalogue where each object is 
surrounded by a mini-DM halo with an inner slope of $ -7/3 $, resulting from 
adiabatic growth of an initial NFW profile (Eq. \ref{eqn:NFW}).

The spike extends from a {\it cut radius} $ r_{cut} $ to the outer
spike radius $ r_{sp} \thickapprox 0.2 \mbox{ } r_h $. The cut radius depends 
on the mass and annihilation cross section of the DM candidate, being 
defined as the radius where the DM density reaches an upper limit due to DM 
annihilations \cite{Bertone_Zentner_Silk}.

Eq. \ref{eqn:annihilation_flux} was used to compute the annihilation flux from 
each mini-spike:
\begin{eqnarray}
\label{eqn:annihilation_flux}
\Phi & = & \int_{E_{thr}}^{m{\chi}} dE \mbox{ }\frac{d\Phi(E)}{dE} \\ 
& = & \int_{E_{thr}}^{m_{\chi}} dE \mbox{ } \frac{1}{2} 
\frac{\sigma v}{m_{\chi}^2} \frac{1}{d^2} \frac{dN_{\gamma}}{dE} 
\int_{r_{cut}}^{r_{sp}} \rho^2_{sp}(r) r^2 dr, \nonumber \\
& = & \Phi_0 \frac{dN_{\gamma}}{dE} 
\left( \frac{\sigma v}{10^{-26} \mbox{cm}^3 s^{-1}} \right)
\left( \frac{m_{\chi}}{100 \mbox{ GeV}} \right)^{-2}
\left( \frac{d}{\mbox{kpc}} \right)^{-2} \nonumber \\
& & \cdot \left( \frac{\rho(r_{sp})}{10^2 \mbox{GeV cm}^{-3}} \right)^2 
\left( \frac{r_{sp}}{\mbox{pc}} \right)^{14/3}
\left( \frac{r_{cut}}{10^{-3} \mbox{pc}} \right)^{-5/3}, \nonumber
\end{eqnarray}

where $ \Phi_0=9 \cdot 10^{-10} \mbox{cm}^{-2}\mbox{s}^{-1} $, $ m_{\chi} $
is the mass of the DM particle and $ \sigma v $ its thermally-averaged
annihilation cross section. $ d $ is the distance of the IMBH from the
detector and $ dN_{\gamma}/dE $ is the number of photons with energy
from $ E $ to $ E+dE $ produced by the annihilation. The differential flux is
then integrated in energy from a lower threshold $ E_{th} $ that depends on 
the experiment considered, up to $ m_{\chi} $.

The differential spectrum $ dN_{\gamma}/dE $ can be written as a sum over
all possible annihilation channels of the number of photons 
$ dN_{\gamma}^a/dE $ produced from the particular channel 
$ \chi \chi \rightarrow a \bar{a} $, weighted by the corresponding branching 
ratio $ B^a=\mbox{BR}(\chi \chi \rightarrow a \bar{a}) $:
\begin{equation}
\frac{dN_{\gamma}}{dE}=\sum_{a}B^a \frac{dN_{\gamma}^a}{dE}.
\end{equation}
 
If DM particles annihilate directly into photons, the spectrum will be
a line at an energy equal to the DM particle mass, but this contribution is usually suppressed at least
for a neutralino-like DM candidate (a branching ratio to primary photons
of $ \thickapprox 10^{-3} $ is used in Ref. \cite{Horiuchi_Ando}).
Otherwise the spectrum will be a continuum, and we can distinguish between
two classes of spectra: in the first one photons are produced from the decay 
of neutral pions 
($ \pi^0 \rightarrow \gamma \gamma $) formed through the fragmentation of  annihilations products such as heavy quarks, gauge or Higgs 
bosons. Many parametrizations are available for these spectra 
$ dN_{\gamma}/dE $ \cite{Fornengo_Pieri_Scopel} 
\cite{Bertone_Servant_Sigl,Fornasa_Taoso_Bertone}, that represent a
suitable approximation for neutralino-like candidates. In the 
right panel of Fig. 7, two possible examples for the 
fragmentations of the neutral pion $ \pi^0 $ are compared 
\cite{Fornengo_Pieri_Scopel,Fornasa_Taoso_Bertone}, in the 
easy-to-consider case of DM annihilating only to a pair of $ b $ quarks
\cite{Bertone_Zentner_Silk}.

The second possibility is inspired from candidates arising in theories with
Universal Extra-Dimensions, where the DM can annihilate to charged leptons
with a large branching ratio (for neutralinos, such channel is severely suppressed). 
Such final states, then, can produce photons through the so-called Final 
State Radiation (FSR). A typical FSR spectrum is plotted in 
Fig. 7 with a 40\% of DM annihilating to light fermions, 20\% to $ \tau $s and 
the remaining in $ b $ quarks. The photon production through light fermions 
is computed analytically \cite{Bergstrom_Bringmann_Eriksson_Gustafsson} while 
the other two channels are described as in the NPD spectrum.

Differences between the two classes of annihilation spectra become
more pronounced above 
$ x \thickapprox 0.6 $, since at low energy the pion decay dominates. The FSR 
spectrum is characterised by a flatter spectrum (the slope in approximately $ -1 $ 
instead of $ -1.5 $ for NPD decay) and by a more abrupt cut-off at 
an energy equal to the DM particle mass. 

The prospects for detecting IMBHs in the MW are summarized in Fig. 8 
where the number of point-like sources (associated to IMBHs) with 
an annihilation flux higher than $ \Phi $, is plotted as a function of $ \Phi $. Compared with the sensitivities of GLAST \cite{GLAST} (a Space-Based 
telescope scheduled to be launched at the beginning of 2008) and EGRET 
\cite{EGRET} (a Space-Based gamma-ray detector, whose data are 
available on Ref. \cite{EGRET_data}) for a $ 5 \sigma $ detection, the most 
optimistic configuration ($ m_{\chi}=100 \mbox{ GeV} $ and 
$ \sigma v=3 \cdot 10^{-26} \mbox{cm}^3 \mbox{s}^{-1}$) leads to almost 
100 (80) detectable sources by GLAST (EGRET) in 1 year.
\begin{figure}
\begin{center}
\includegraphics[width=8cm]{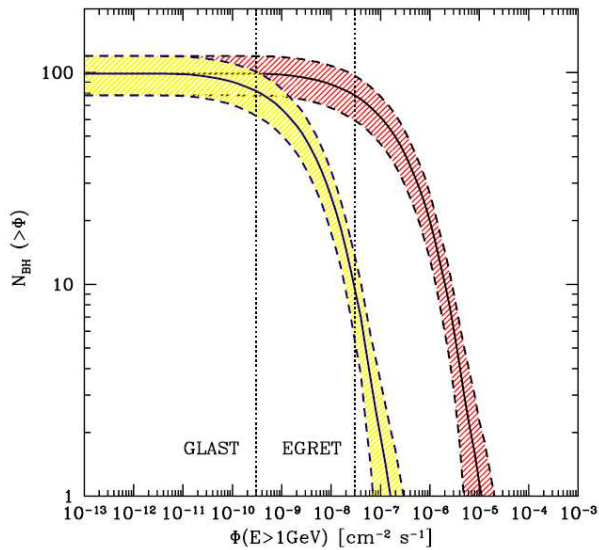}
\label{fig:visible_IMBHs}
\caption{IMBHs integrated luminosity function, i.e. number of BHs producing a gamma-ray flux larger than a given flux, as a function of the flux, for scenario B and a NPD spectrum. The upper (lower) line corresponds to $ m_{\chi}=100 \mbox{ GeV} $, $ \sigma v=3 \cdot 10^{-26} \mbox{cm}^3 \mbox{s}^{-1} $ ($ m_{\chi}=1 \mbox{ TeV} $, $ \sigma v=10^{-29} \mbox{cm}^3 \mbox{s}^{-1} $). For each curve we also show the 1$ \sigma $ scatter among different realization of the MW DM halo. The figure can be interpreted as the number of IMBHs that can be detected from experiments with a point-source sensitivity $ \Phi $ (above 1 GeV), as a function of $ \Phi $. We show for comparison the $ 5 \sigma $ point-source sensitivity above 1 GeV of EGRET and GLAST (1 year). Figure taken from Bertone, Zentner and Silk (2005).}
\end{center}
\end{figure}

The prospects for Indirect DM detection usually strongly depend on the 
particle physics parameters. For IMBHs, however, the $ \sigma v/m_{\chi}^2 $ 
dependence of the flux in Eq. \ref{eqn:annihilation_flux} is modified by the 
implicit dependence of the cut radius on the mass $ m_{\chi} $ and the 
annihilation cross section $ \sigma v $, so that at the end
$ d\Phi/dE \propto (\sigma v)^{2/7} m_{\chi}^{-9/7} $.

In order to discriminate IMBHs from other extragalactic sources, one can 
perform a maximum likelihood analysis of the annihilation spectrum or look
for a class of sources with an identical cut-off, at the mass of the DM 
particle \cite{Bertone_Zentner_Silk}. Alternatively, one may look at the 
Andromeda galaxy M31 \cite{Fornasa_Taoso_Bertone}, which is located 
$ \thickapprox 780 \mbox{ kpc} $ away from us, and where a population of 
$ 65 \pm 15 $ objects (in scenario B) is predicted. A fraction of these BHs 
($ 17 \pm 6 $ for $ m_{\chi}=150 \mbox{ GeV} $ with a NPD spectrum) can be
detected with GLAST. In this case, the DM signature will be the detection of
up to 20 point-like, bright, gamma-ray sources in a $ 3^{\circ} $ circle 
around the Andromeda center (see Fig. 9).
 
\begin{figure}
\begin{center}
\includegraphics[width=8cm]{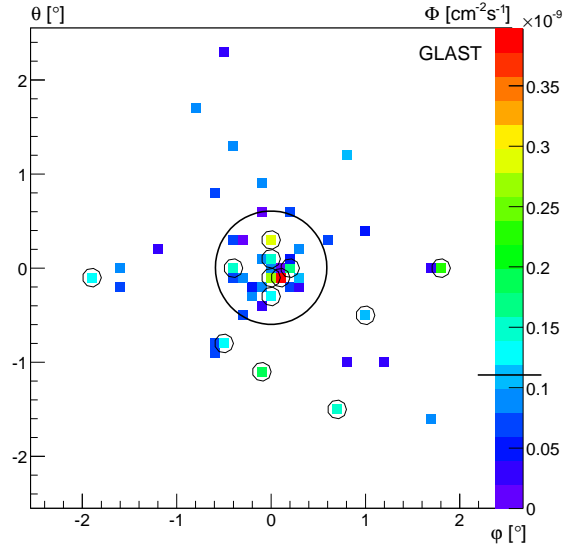}
\label{fig:Fornasa_Taoso_Bertone}
\caption{Map of the gamma-ray flux, in units of photons $ \mbox{cm}^{-2} \mbox{s}^{-1} $, from DM annihilations around IMBHs in M31, relative to one random realization of IMBHs in M31. The size of the bins is $ 0.1^{\circ} $ and the energy threshold is 4 GeV, as appropriate for GLAST. The big circle shows, for comparison, the M31 scale radius $ r_s $, and the small circles highlight IMBHs within the reach of GLAST ($ 5 \sigma $ detection in 2 months). Figure taken from Fornasa, Taoso and Bertone (2007).}
\end{center}
\end{figure}

\subsection{Contribution of SMBHs and IMBHs to the Gamma-ray Background}
\label{sec:EGB}
EGRET data have shown the existence of an Extragalactive diffuse Gamma-ray 
Background (EGB) in the GeV range \cite{Sreekumar_et_al,Strong_Moskalenko_Reimer}, initially interpreted as photons from an 
unresolved population of blazars \cite{Narumoto_Totani}. Today such 
astrophysical sources are believed to play a significant role but they can 
hardly account for the entire background, while DM spikes and mini-spikes could
in principle strongly contribute to the flux.

Ullio {\it et al.} \cite{Ullio_Bergstrom_Edsjo_Lacey} estimated the
contribution of DM annihilations from halos without central BHs to the 
EGB. The total number of photons that contributes to the EGB from redshift 
$ z $ is in this case:  

\begin{eqnarray}
\label{eqn:annihilation_redshift}
d \mathcal{N}_{\gamma} & = & \int dM (1+z)^3 \frac{dn}{dM}(M,z)
\frac{\sigma v}{2m_{\chi}^2} \frac{dN_{\gamma}(E)}{dE} \cdot \\
& & \int d {\bf r} \rho^2({\bf r},M,z) e^{-\tau(E_0,z)} \frac{dVdA}{4\pi(R_0r)^2} dE_0 dt_0 \nonumber,
\end{eqnarray}
where $ dn/dM $ is the comoving number density of DM halos with mass $ M $ at
redshift $ z $ and it is computed using the Press-Schechter formalism 
\cite{Press_Schechter,Ando}. 
The $ (1+z)^3 $ factor converts comoving to proper density, while 
$ dN_{\gamma}/dE $ is the usual annihilation spectrum described in the last 
Section. $ \rho $ is the DM density and the factor $ e^{-\tau(E_0,z)} $ 
takes into account the gamma ray absorption due to pair production with the 
ExtraGalactic Background Light in the infrared or optical band 
($ \tau \thickapprox (E_0/10 \mbox{ GeV})^{0.8} (z/3.3) $). $ E_0 $ is the 
energy today so that $ dEdt=dE_0(1+z) dt_0(1+z)^{-1} $ and $ R_0 $ is the today
scale factor. 

Each halo is described by a NFW (Eq. \ref{eqn:NFW}) or M99 (Eq. \ref{eqn:M99})
profile,
whose concentration parameter $ c$ ($ c=r_{vir}/r_s $) depends on redshift
as in Eke, Navarro and Steinmetz \cite{Eke_Navarro_Steinmetz}.
In order to reproduce the EGRET diffuse flux, a large boost factor would be 
required, that, if applied to the Galactic center, would grossly exceed the
EGRET measurement of the gamma ray flux from Sgr A$ ^{\ast} $ \cite{Ando}.

The presence of a spike would only marginally affect the signal from the Galactic 
center, because even if the spike was present in the past for the SMBH of the 
MW, it would already be destroyed. However,for the EGB spikes are indeed
important, since the background receives contributions from halos at high 
redshift, at a time when astrophysical and particle physics effects did not 
have time yet to damp DM overdensities \cite{Ahn_Bertone_Merritt_Zhang}. 

Ahn {\it et al.} \cite{Ahn_Bertone_Merritt_Zhang} included spikes around SMBHs 
in the computation made by Ullio {\it et al.}. The authors used the 
phenomelogical relation between the mass of the BH $ M_{\bullet} $ and the 
mass of the hosting halo $ M_{h} $ \cite{Ferrarese}: 
$ M_{\bullet}/10^8 M_{\odot}=a(M_h/10^{12}M_{\odot})^b $, where $ M_h $ is 
considered at $ z=0 $ and different values for $ (a,b) $ are used
\cite{Ahn_Bertone_Merritt_Zhang}.

For each halo at $ z=0 $, a SMBH with mass $ M_{\bullet} $ (related to
$ M_h $) is posed at the redshift of BH formation $ z=z_{\bullet} $ and the 
halo is evolved till now (assuming that $ M_h $ remains unchanged). The spikes 
form and then decay in a self-similar way (see Section \ref{sec:CREST}, 
$ \rho(r,t) \thickapprox \rho(r,0) \kappa= \rho(r,0) e^{-\tau/2} $), so that
the evolution of the spike is described by
\begin{equation}
\rho_{sp}(r,t)=\rho_{sp,0} \kappa^{\epsilon} 
\left( \frac{r}{r_{s,0}}\right)^{-\gamma_{sp}},
\end{equation}

with $ \gamma $ and $ \gamma_{sp} $ are the slopes of the initial and final
profile, $ \epsilon=(\gamma_{sp}-\gamma)^{-1} $ and $ \tau $ is the time
since spike formation in units of the heating time $ t_{heat} $:
\begin{equation}
t_{heat}=1.25 \mbox{ yr} \cdot 
\left( \frac{M_{\bullet}}{3 \cdot 10^6 M_{\odot}} \right)^{1/2}
\left( \frac{r_h}{2 \mbox{ kpc}} \right)^{3/2}
\left( \frac{m_{\star}}{M_{\odot}} \right)^{-1}
\left( \frac{\ln \Lambda}{15} \right)^{-1}.
\end{equation}

If we specify a NFW profile for the DM density before the spike formation, 
the complete DM distribution can be written as follows:
\begin{equation}
\left\{
\begin{array}{lcl}
\rho(r)_{NFW} & & (z > z_{\bullet}) \\
\rho(r)_{NFW} & & (z \leq z_{\bullet}, r > r_{s,0}) \\
\rho(r)_{NFW}+\rho_{sp}(r,t) \thickapprox \rho_{sp}(r,t) & &
(z \leq z_{\bullet}, r \leq r_{s,0}) \\
\end{array} 
\right. .
\end{equation}

For a DM particle of 100 GeV the contribution to the EGB is summarized in 
the left panel of Fig. 10, that shows an enhancement of more than an order of 
magnitude with respect to the case without SMBHs. The main contribution comes 
from low-energies, due to the annihilations at high redshift in the 
still-present spikes. The energy of the produced photons is then redshifted 
till now. 

\begin{figure}
\begin{center}
\includegraphics[width=5.9cm]{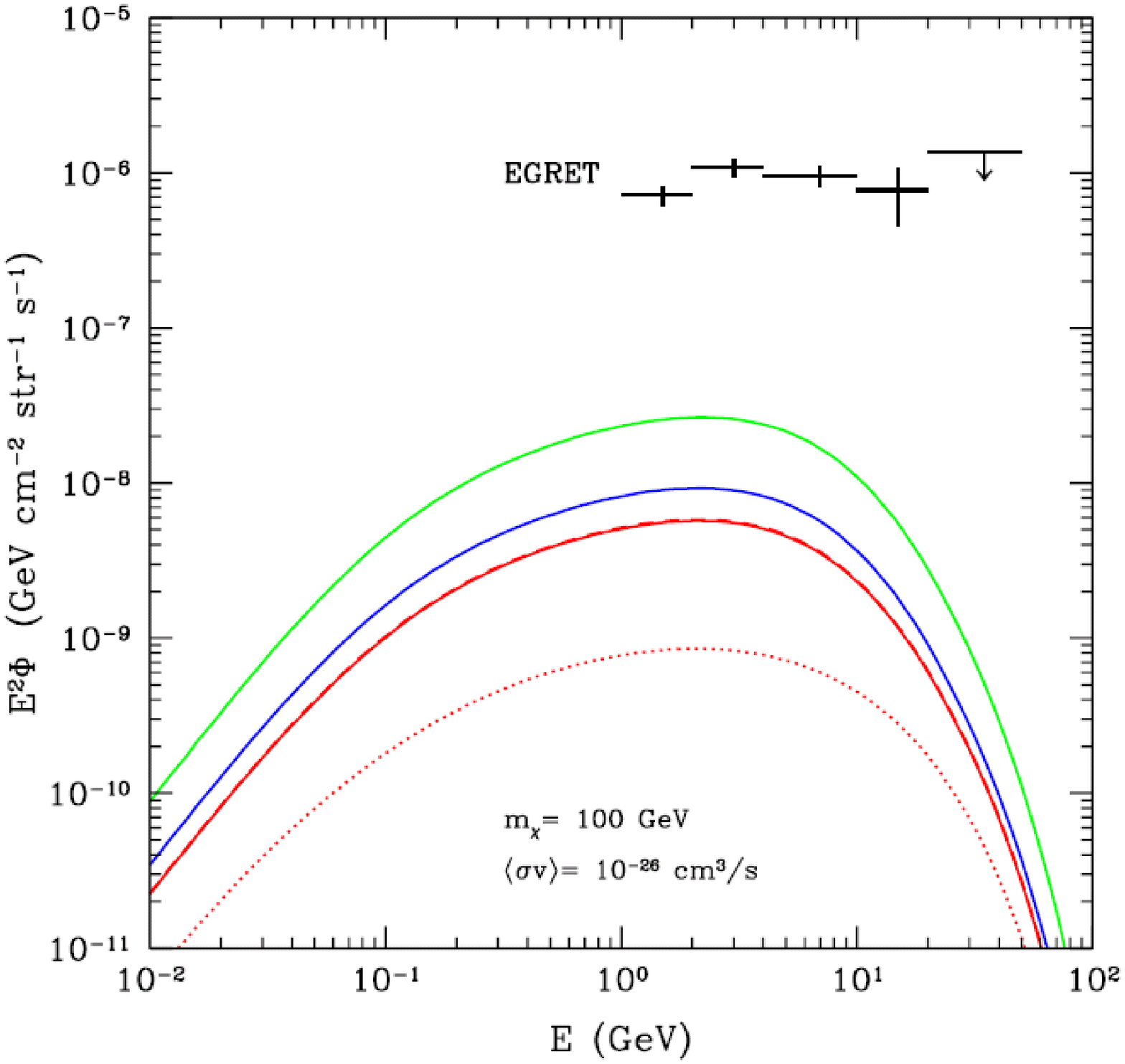}
\includegraphics[width=5.5cm]{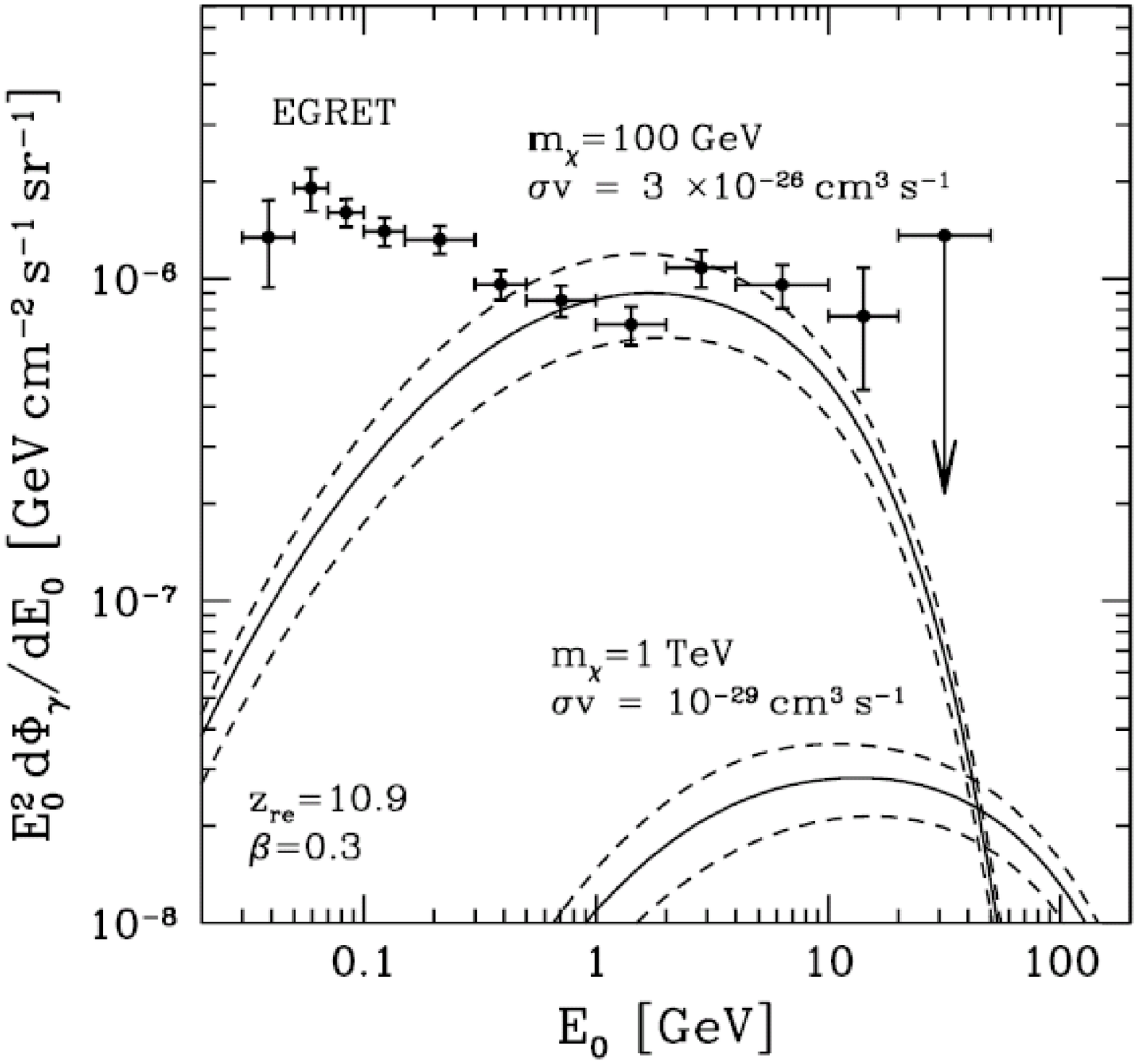}
\label{fig:EGB_SMBHs}
\caption{{\it Left:} Gamma-ray background produced by DM annihilations in DM halos with spikes (solid lines), compared to the halo-only contribution (dotted). The EGRET diffuse flux limits are shown for comparison. Three choices for the parameters in the relation between tha mass of the BH and the mass of the halo are used: $ (a,b)=(0.027,1.82) $ in red, $ (0.10,1.65) $ in blue and $ (0.67,1.82) $ in green. The DM parameters adopted are $ m_{\chi}=100 \mbox{ GeV} $, $ \sigma v=10^{-26} \mbox{cm}^3 \mbox{s}^{-1} $. Figure taken from Ahn, Bertone, Merritt and Zhang (2007) {\it Right:} Contribution to the EGB from scenario A and scenario B IMBH mini-spikes. Also shown are the EGRET data and predictions for the case of standardhalos without spikes and mini-spikes. The $ 1 \sigma $ scatter in the number of IMBHs is shown for scenario A. For scenario B, the $ 1 \sigma $ scatter for the number of IMBHs and for the upper limit in the redshift integration are shown combined. Results are shown for $ m_{\chi}=100 \mbox{ GeV} $ and $ \sigma v = 3 \cdot 10^{-26} \mbox{cm}^3 \mbox{s}^{-1} $. Figure taken from Horiuchi and Ando (2006).}
\end{center} 
\end{figure}

The EGB from DM annihilation in mini-spikes around IMBHs has been calculated
in Ref. \cite{Horiuchi_Ando}.
The results are shown in the right panel of Fig. 10 for a neutralino
of $ 100 \mbox{ GeV} $ and both the formation scenarios. As before, the main 
contribution is at low energies, but now the predictions from scenario B are 
able to account for all the EGB, and the same is true if one consider the line 
spectrum.
This model is sensitive to the average of the halos properties, whereas, in 
the case of annihilation e.g. from the Galactic Center, one has to deal with 
a single realization that may differ significantly from the average, so that 
the study of gamma ray background and the way DM contributes is, for Indirect 
searches, an interesting alternative to the study of unidentified sources.
\begin{figure}[t]
\hspace{1.0in}
\includegraphics[width=0.8\textwidth]{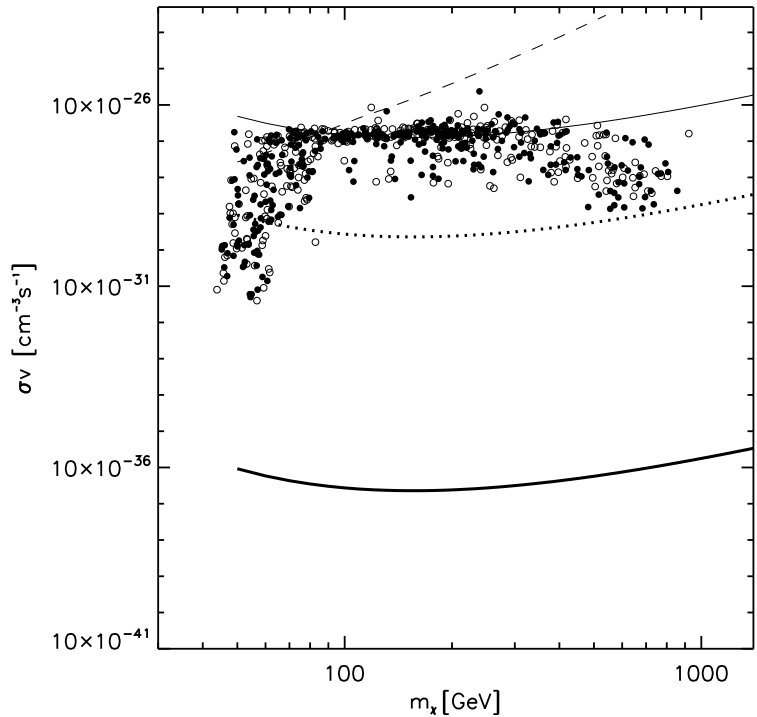}
\label{exclusion}
\caption{Exclusion plot in the DM mass vs. annihilation cross section plane. Above the thick (dotted) solid line all the statistical realizations of a MW-like halo in Bertone, Zentner and Silk (2005) contain at least one IMBH with an annihilation flux larger than the estimated point source sensitivity of GLAST (EGRET) in 2 months. Above the thin solid line all the statistical realizations of the Andromeda galaxy in Fornasa, Taoso and Bertone (2007) have at least one IMBH with an annihilation flux larger than $ 10^{-10} \mbox{cm}^{-2} \mbox{s}^{-1} $. The dotted line indicates the region above which the contribution from DM annihilation around IMBHs at high redshift can account for the EGB as measured by EGRET. The empty (full) dots represent SUSY models (obtained with DarkSusy) where the neutralino relic density is within 3 (5)$ \sigma $ from the WMAP-measured value.}
\end{figure}

\section{Conclusions}
\label{sec:six}
We have reviewed the consequences of the growth and evolution of BHs 
on the distribution of stars and DM around 
them. Initially, the BH mass was considered to be fixed in time, with 
particles evolving around it under the influence of its gravitational 
potential and their gravitational collisions. A steady-state solution is 
predicted for stars only for collisional nuclei: it is a zero-flux solution 
characterized by a central cusp with a slope of $ -7/4 $ \cite{Bahcall_Wolf} 
within $ 0.2 \mbox{ } r_h $. DM particles, being collisionless, are not 
affected by self gravitational interactions but by encounters with stars 
\cite{Merritt_Harfst_Bertone}: in the same time that the $ -7/4 $ cusp forms 
for stars, the DM distribution is rearranged in a profile with a central 
slope of $ -3/2 $, that will decay due to heating.
Other effects can significantly reduce collisional overdensities, e.g. both
stellar and DM cusps react to the presence of a BH binary 
\cite{Merritt_binaries} and the stellar number density can be influenced by 
loss-cone dynamics 
\cite{Merritt}.

In the case of adiabatic growth of a BH, instead, the final DM and stellar 
overdensities are much steeper, reaching slope as large as $ -2.25 $ 
\cite{Quinlan_Hernquist_Sigurdsson,Gondolo_Silk}. But again, the 
evolution of the host galaxy drives a suppression of the spike, due to 
cumulative mergers \cite{Merritt_Milosavljevic_Verde_Jimenez} and possible 
off-center BH formation \cite{Ullio_Zhao_Kamionkowski}.
Even if DM overdensities around BHs may seem to be very promising 
for Indirect detection, the presence of spikes is probably not realistic 
\cite{Ullio_Zhao_Kamionkowski}. Alternative strategies include the study of
DM contribution to the EGB and the search for IMBHs. In the first case, the 
gamma-ray background receives contribution from BHs at high redshift, when
spikes and mini-spikes were already present, but damping mechanisms were not
effective yet. DM annihilations from spikes around SMBHs at high redshift 
contribute significantly to the EGB \cite{Ahn_Bertone_Merritt_Zhang}. A 
population of IMBHs is then able to reproduce the value detected by EGRET, even
for standard assumptions for the initial DM profile \cite{Horiuchi_Ando}.

Mini-spikes around IMBHs can effectively be searched for with GLAST and can 
be identified with a careful spectral analysis of the signal 
\cite{Bertone_Zentner_Silk}. They can also been searched for in the Andromeda
galaxy, distributed in a quite characteristic fashion that leaves no room for
alternative astrophysical interpretations \cite{Fornasa_Taoso_Bertone}.

We show in Fig. 11 the constraining power of the different detection 
strategies discussed above:
\begin{itemize}
\item the dotted and solid lines are relative to the population of IMBHs in our Galaxy 
\cite{Bertone_Zentner_Silk}. For each value of the mass of 
the DM candidate, we estimate the value of 
$ \sigma v $ for which all the aformentioned realizations of the Milky Way 
have {\it at least one} IMBH that should have been detected by EGRET 
at $ 5 \sigma $, as discussed in 
Ref. \cite{Bertone_Zentner_Silk}. In other words, if none of the EGRET
unidentified sources can be interpreted as IMBH, a model with a combination
$ (m_{\chi},\sigma v) $ above the dotted line is compatible with
the data with a probability less than 0.5\%. The solid line shows 
the prospectesd for detection with GLAST
\item the thin solid line is relative to the population of IMBHs in M31
\cite{Fornasa_Taoso_Bertone}. For each value of $ m_{\chi} $ we determined the
value of $ \sigma v $ for which all the 200 realizations of M31 have at least
one IMBH with an annihilation flux larger than the $ 5 \sigma $ sensitivity of 
GLAST (parametrized as in Ref. \cite{Fornasa_Taoso_Bertone}), for an
exposure of 2 months, and an energy threshold of 4 GeV. Null searches of IMBHs
in Andromeda will exclude the region above the thin line
\item the dashed line is relative to the contribution of IMBHs to 
the EGB \cite{Horiuchi_Ando}. Each point above the line corresponds
to a DM candidate for which annihilations in IMBHs mini-spikes at high 
redshift exceeds the EGB values as measured by EGRET in 
Ref. \cite{Sreekumar_et_al}.
\item the empty dots represent Supersymmetric Models (generated
with DarkSusy \cite{DarkSusy}) for which the neutralino can account 
for all the DM, being within $ 3\sigma $ from the WMAP determination 
of the DM relic density
\cite{WMAP} 
\item the full dots represent SUSY Models 
within $ 5 \sigma $ from WMAP 
\end{itemize}

GLAST can thus probe regions of the parameter space at very low cross
section, so that IMBHs can be soon discovered or ruled 
out.  If none of these sources is detected by GLAST, it is very unlikely than such scenario will survive, at least for
a WIMP DM candidate.

The gamma-ray flux from DM annihilations in spikes and mini-spikes has
been widely discussed in literature (Refs. 
\cite{Ullio_Zhao_Kamionkowski,Gondolo_Silk,Merritt_Milosavljevic_Verde_Jimenez,Merritt_Harfst_Bertone,Bertone_Merritt,Zaharijas_Hooper,Pieri_Branchini}, 
just to name a few) and many experimental 
collaborations are actively working to improve their discovery potential. 
But so far, the gravitational effects of BHs have only been observed through their  impact on 
stellar populations, since there is not a detected signal currently and clearly 
interpretated as DM annihilation.
Photons from DM annihilations could have been detected in the past as 
subdominant contributions from gamma-ray sources \cite{Aharonian_et_al} but 
the main challenge remains to disentagle such a contribution 
from other photon production mechanisms.

Instead, the effect of a central BH on the stellar population has been
detected for the MW \cite{Genzel_et_al,Schodel_et_al} and such
interpretation is proposed also for other galaxies of the Local Group 
\cite{Lauer,Young_et_al}, encouraging further attempts of detecting BH 
effects on DM and, eventually, of detecting DM indirectly through its 
annihilations products.

\begin{acknowledgments}
We thank David Merritt for many useful comments and suggestions.
\end{acknowledgments}

\end{document}